\documentclass[manuscript, screen]{acmart}
\usepackage{booktabs} 

\usepackage[ruled]{algorithm2e} 

\newenvironment{myitemize}[1][]{
\begin{list}{$\bullet$}
    {
     \setlength{\leftmargin}{2mm}     %
     \setlength{\parsep}{1mm}         %
     \setlength{\topsep}{0mm}         %
     \setlength{\itemsep}{0mm}        %
     \setlength{\labelsep}{1.5mm}     %
     \setlength{\itemindent}{0mm}    %
     \setlength{\listparindent}{5mm} %
    }}
{\end{list}}

\usepackage{multirow}
\usepackage{multicol}
\usepackage[flushleft]{threeparttable}
\newcommand{\tabincell}[2]{\begin{tabular}{@{}#1@{}}#2\end{tabular}}
\usepackage{lscape}
\usepackage{enumerate}

\acmJournal{CSUR}

\setcopyright{acmcopyright}

\acmDOI{0000001.0000001}

\begin{document}
\title{ A Survey on  Food Computing}

\author{Weiqing Min}
\affiliation{%
  \institution{Key Lab of Intelligent Information Processing, Institute of Computing Technology, CAS}
  \city{Beijing}
  \country{China}}
\email{minweiqing@ict.ac.cn}
\author{Shuqiang Jiang}
\affiliation{%
  \institution{Key Lab of Intelligent Information Processing, Institute of Computing Technology, CAS}
  \city{Beijing}
  \country{China}}
\email{sqjiang@ict.ac.cn}
\author{Linhu Liu}
\affiliation{%
  \institution{Key Lab of Intelligent Information Processing, Institute of Computing Technology, CAS}
  \city{Beijing}
  \country{China}}
\email{linhu.liu@vipl.ict.ac.cn}
\author{Yong Rui}
\affiliation{%
 \institution{Lenovo Group}
 \streetaddress{No. 6, Shangdi West Road}
 \city{Beijing}
 \country{China}}
\email{yongrui@lenovo.com}
\author{Ramesh Jain}
\affiliation{%
  \institution{Department of Computer Science, University of California}
  \city{Irvine}
  \state{CA}
  \country{USA}
}
\email{jain@ics.uci.edu}

\begin{abstract}
Food is very essential for human life and it is fundamental to the human experience. Food-related study may support multifarious applications and services, such as guiding the human behavior, improving the human health and understanding the culinary culture. With the rapid development of social networks, mobile networks, and Internet of Things (IoT), people commonly upload, share, and record food images, recipes, cooking videos, and food diaries, leading to large-scale food data. Large-scale food data offers rich knowledge about food and can help tackle many central issues of human society. Therefore, it is time to group several disparate issues related to  food computing. Food computing acquires and analyzes heterogenous food data from disparate sources for perception, recognition, retrieval, recommendation,  and monitoring of food.  In food computing,  computational approaches are applied to address food related issues in medicine, biology, gastronomy and agronomy. Both large-scale food data and recent breakthroughs in computer science are transforming the way we analyze food data. Therefore, vast amounts of  work has been conducted in the food area, targeting different food-oriented tasks and applications. However, there are very few systematic reviews, which shape this area well and provide a comprehensive and in-depth summary of current efforts or detail open problems in this area. In this paper, we formalize food computing and present such a comprehensive overview of various emerging concepts,  methods, and tasks.  We summarize key challenges and future directions ahead for food computing. This is the first comprehensive survey that targets the study of computing technology for the food area and also offers a collection of research studies and technologies to benefit researchers and practitioners working in different food-related fields.
\end{abstract}

%
%


\maketitle

\section{Introduction}
Food has a  profound impact on human life, health and wellbeing \cite{Nordstr2013Food,Achananuparp2018Does}. An increasing  amount of people  is becoming  overweight or obese. According to WHO, there are more than 1.9 billion adults aged 18 or over  with overweight, where more than 650 million ones are obese. The worldwide prevalence of obesity in 2016 is  nearly three times that of 1975\footnote{http://www.who.int/news-room/fact-sheets/detail/obesity-and-overweight}.  Overweight and obesity have been found to be one of major risk factors for various chronic diseases, such as  diabetes and cardiovascular diseases\footnote{ http://www.who.int/mediacentre/factsheets/fs311/en/index.html}. For example, it is estimated that 415 million people suffers from diabetes worldwide in 2015\footnote{http://www.diabetesatlas.org/}. One important reason is that many  generally maintain an excessive unhealthy lifestyle and bad dietary habits \cite{Ng2014Global}, such as the increased intake of  food with high energy and high fat. In addition, food is much more than a tool of survival. It plays an important role in defining our identity, social status, religious significance and culture \cite{Harris-G2E-AA1985,Khanna-FC-EFN2009}. Just as Jean Anthelme Brillat-Savarin said, ``tell me what you eat, and I will tell you who you are".  Furthermore, how we cook it and  how we eat it are also factors profoundly touched by our individual cultural inheritance. For these reasons, food-related study \cite{Canetti2002Food,Ahn-FNFP-SciRe2011,bucher2013fruit,Sajadmanesh-KC-arXiv2016,Chung2017A} has always been  a  hotspot and received extensive attention from various fields.

In the earlier years,  food-related study has been  conducted from different aspects, such as food choice~\cite{Nestle1998Behavioral}, food perception~\cite{S2003Effect}, food consumption~\cite{Pauly1986A}, food safety~\cite{Chen2001Food} and food culture~\cite{Harris-G2E-AA1985}. However, these studies are conducted using traditional approaches before the web revolutionized research in many areas. In addition, most methods use a small-scale data, such as questionnaires, cookbooks and  recipes. Nowadays, the fast development of various networks, such as social networks, mobile networks and Internet of Things (IoT)  allows users to easily  share  food images, recipes, cooking videos or record food diary via these networks, leading to large-scale food dataset. These food data implies rich knowledge and thus can provide great opportunities for food-related study, such as  discovering principles of  food perception~\cite{mouritsen2017data}, analyzing culinary habits \cite{Sajadmanesh-KC-arXiv2016} and monitoring the diet  \cite{Chung2017A}.  In addition, various new data analysis methods in network analysis, computer vision, machine learning and data mining  are proposed.  Recent breakthroughs in Artificial Intelligence (AI), such as deep learning \cite{LeCun-DL-Nature2015,Jordan255} have further fueled the interest in  large-scale food-oriented study \cite{Kawano2014Food,Hassanne-FIRDCN-MM2016,Pandey2017FoodNet,Chen2017Counting} for their superiority in learning  representations from various types of signals.

Taking these factors into consideration, we come up with a vision of  food computing, which aims to apply   heterogeneous food data collected from different data sources to  various applications in different fields, such as the human behavior \cite{Kolata1982Food}, health \cite{Harvey2017Exploiting}, agriculture \cite{Hern-SOC-ICTI2017}, culture \cite{Ahn-FNFP-SciRe2011}, medicine \cite{Batt2007Food}, food security \cite{Barrett2010Measuring} and food science \cite{Ofli-Saki-WWW2017}. To our knowledge, \cite{Harper2015OpenAG} first proposed the term food computing, which is particularly used  in the agricultural field. However, they didn't give  clear definition. In a broad sense, we think that food computing focuses on food-related study via  computer science, and it is an interdisciplinary field. Consequently, there are many   open questions to answer. For example, what are the core research problems of  food computing?  What are the key methodologies for  food computing?  What are  representative applications in this domain? What are  challenges and potential directions for this research field?

To answer these questions, we formally define  food computing in this article and introduce its general framework,  tasks and applications. Some food-related surveys have been done. For example, \cite{Knez2015Food} gave a  survey on mobile food recognition and  nine  recognition systems are introduced based on their system architecture. \cite{Trattner2017Food} provided a summary of   food recommender systems. \cite{BVR-AFM-JHMI2017} presented a variety of methodologies and resources on automatic food monitoring and diet management system.  However, to the best of our knowledge, there are very few systematic reviews, which  shape this area  well and  provide a comprehensive and in-depth summary of current efforts, challenges or future directions  in the area. This survey seeks to provide such a comprehensive summary of current research on food computing  to identify open problems and  point out future directions. It aims to build the connection between computer science and food-related fields, serving as a good reference for developing  food computing techniques and applications for various food-related fields. To this end, about 300 studies are shortlisted and classified in this survey.

This survey is organized as follows: Section \ref{section_FC} first presents the concept and  framework of  food computing. Section \ref{section_FDAA} introduces  food data acquisition and analysis, where different types of food datasets are summarized and compared.  We present its representative applications  in Section \ref{section_AFC}. Main tasks in  food computing are reviewed in Section \ref{section_FCT}. Section \ref{section_CHA} and Section \ref{section_FD}  discuss its challenges and prominent open research issues, respectively. We finally  concludes the article in Section \ref{section_CONCLU}.


\section{Food Computing}\label{section_FC}
\begin{figure*}
\centering
 \includegraphics[width=0.70\textwidth]{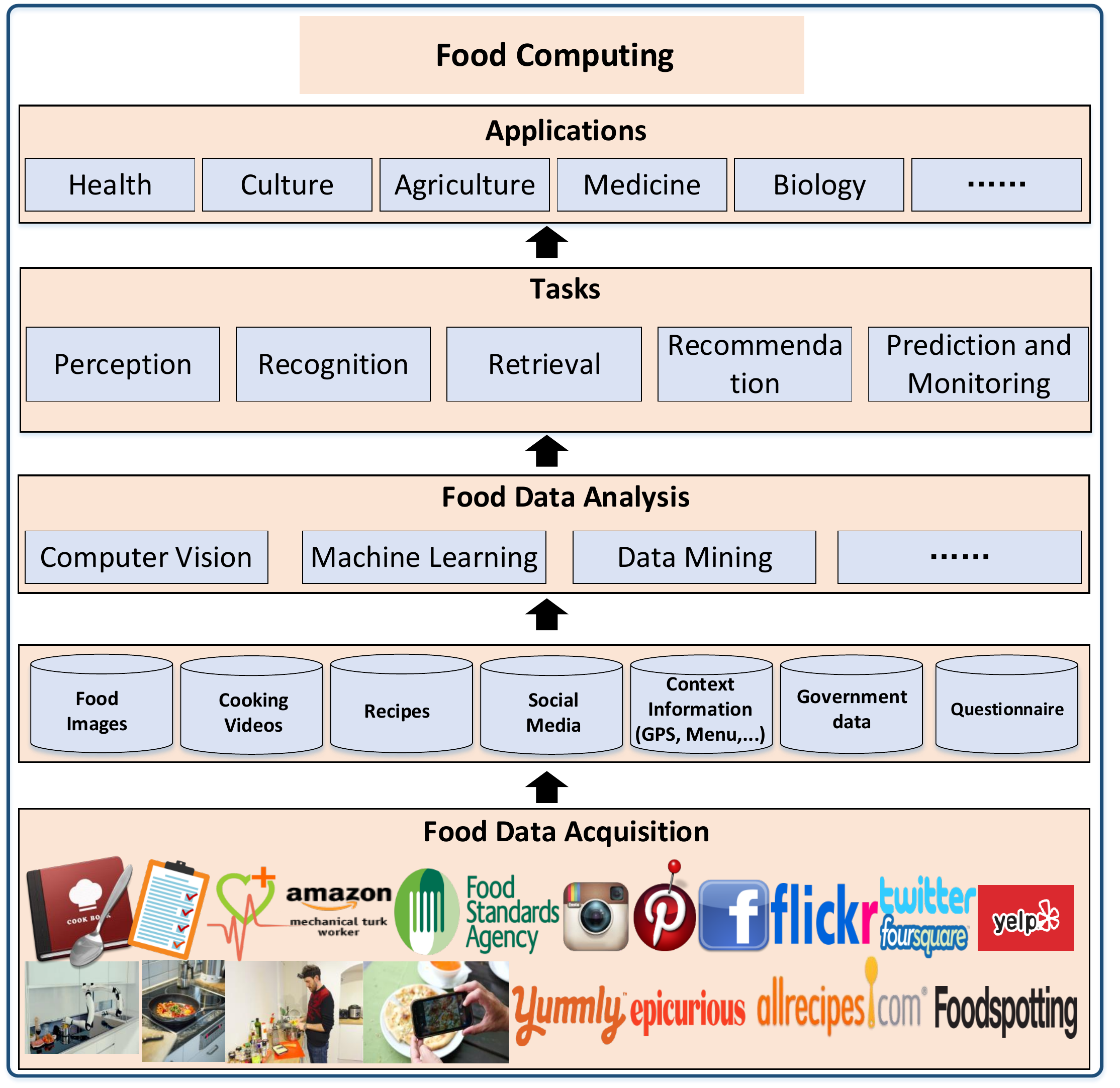}
\caption{An overview of  food computing.}
\label{MFC}       
\end{figure*}
{Food computing} mainly utilizes the methods from  computer science for food-related study.  It involves the acquisition and analysis of  food data with different modalities (e.g., food images, food logs, recipe, taste and smell) from different data sources (e.g., the social network, IoT and recipe-sharing websites). Such analysis resorts to computer vision, machine learning, data mining and other advanced technologies to connect food and human for supporting  human-centric  services, such as improving the human health, guiding the human behavior  and understanding the human culture. It  is  a typically  multidisciplinary field, where computer science meets conventional food-related fields, like food science, medicine, biology, agronomy, sociology and gastronomy. Therefore, besides computer science, food computing also borrows theories and methods from other disciplines, such as neuroscience, cognitive science and chemistry. As shown in Figure~\ref{MFC}, food computing mainly consists of five basic tasks, from perception, recognition, retrieval, recommendation to prediction and monitoring. It further enables various applications for many fields, such as health, culture, agriculture and medicine.

\textit{Food computing applies computational approaches for acquiring and analyzing heterogenous food data from disparate sources for perception, recognition, retrieval, recommendation and monitoring of food to address food related issues in health, biology, gastronomy and agronomy.}

Figure~\ref{MFC} shows its general framework. One  important goal of food computing is to provide various human-centric services. Therefore, the first step is to collect human-produced food data. We can acquire  food data with differen types from various data sources, such as cookbooks, social networks, various sensors, IoT and  recipe-sharing websites. In addition, there are also other specific food dataset available, such as the odor threshold database and the Volatile Compounds in food database. Based on these food data, we utilize machine learning, computing vision, data mining and other technologies for  food data analysis. After  that, we can conduct a series of  food computing tasks, where food-oriented perception, recognition, retrieval, recommendation, prediction and monitoring together constitute  main tasks of  food computing. The flavor and sensory perception of food can govern our choice of food and affect how much we eat or drink. Food perception is multi-modal, including visual information, tastes, smells and tactile sensations. Recognition is one basic task and it  is mainly to predict  food items such as the category or ingredients from  food images. Food-oriented retrieval involves single-modality based retrieval (such as visual food retrieval and recipe retrieval) and cross-modal retrieval, which receives more attention for its applications such as retrieving the recipes from food images. Food-oriented recommendation can not only recommend the food  people might want to eat, but also provide them with a healthier diet. Food recommendation involves more complex and multi-faceted information. Therefore, it is different from other types of recommendations. Prediction and monitoring are mainly conducted based on the social media, such as monitoring public health.

Furthermore, different tasks are not independent but closely intertwined and mutually dependent. For example, the recognized results can further support  the retrieval and recommendation tasks. More and more works also resort to  the recognition technology for food perception \cite{Ofli-Saki-WWW2017}. When the categories of food images are huge, retrieval-based method can also be used for food recognition. Prediction from the social media can also be helpful for  the  recommendation task. For example,user's food preference predicted from social media will be  important information for personalized food recommendation.

\section{Food Data Acquisition and Analysis}\label{section_FDAA}
In this section, we introduce frequently used data sources in food computing and briefly give the summary and comparison on existing food datasets with different types.

Benefitting from the development of the internet and various smart devices, an amount of research work focuses on studying the food perception, pattern mining and human behavior via various data-driven methods  \cite{mouritsen2017data}. For example, in order to analyze user's eating habit for his/her dietary assessment, we should acquire his/her food log data for further analysis.  Through  the analysis of these food data, we can discover some general principles that may underlie food perception  and  diverse culinary practice. Therefore, the first step of food computing involves the acquisition and collection of food data. Particularly, we summarize various data sources into  five  types: (1) official organizations and experts; (2)  recipe-sharing websites; (3) social media; (4) IoT devices and (5) crowd-sourcing platforms.

In the early years, researchers mainly obtain  food data from  official organizations or  experts to conduct food-related study. For example, \cite{Sherman1999Darwinian} analyzed the recipes  to find the reason that humans use spices using 93 traditional cookbooks from 36 counties. In order to calculate the food calorie, they should search its  energy in the nutrition table provided by official organizations, e.g., United States Department of Agriculture (USDA)\footnote{https://ndb.nal.usda.gov/ndb/} and BLS\footnote{https://www.blsdb.de}. \cite{Martin2012Validity}  resorted to  nutrition experts  to label food items. These  data acquisition methods are generally time-consuming, laborious and hard to achieve the large-scale.
\begin{figure*}
\centering
 \includegraphics[width=0.70\textwidth]{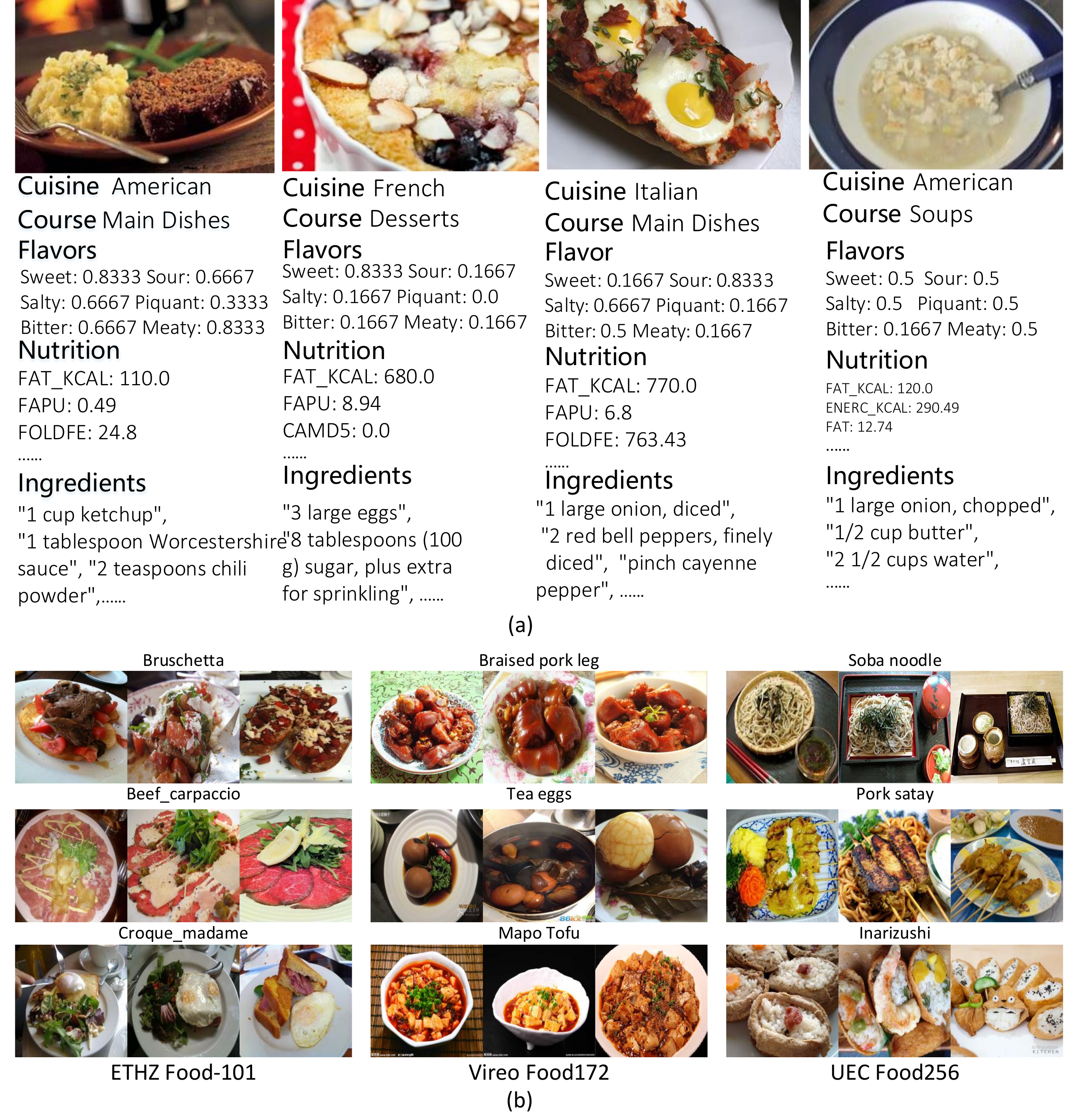}
\caption{(a) Some recipes from Yummly (b) Food images from two datasets: ETHZ Food-101 ~\cite{Bossard-Food101-ECCV2014} and UEC Food256 ~\cite{kawano2014automatic}}
\label{dataset_samples}       
\end{figure*}

The  proliferation of recipe-sharing websites has resulted in  huge online food data collections. Many recipe-sharing websites such as Yummly\footnote{http://www.yummly.com/}, Meishijie\footnote{http://www.meishij.net/}, foodspotting.com and Allrecipes.com have emerged over the last several years. Besides basic information, such as the list of ingredients, these recipes are associated with rich modality and attribute information. Figure \ref{dataset_samples}~(a) shows some examples from Yummly. Each recipe includes a list of ingredients, the food image, cuisine category, course, flavor and macronutrient composition.  Such  recipe data with rich types can be exploited to answer various food related questions. For example, \cite{Ahn-FNFP-SciRe2011}  used  recipes from three  repositories (epicurious.com, menupan.com and Allrecipes) to analyze  patterns of ingredient combination based on flavor compounds from different regions. \cite{Bossard-Food101-ECCV2014} constructed the first large-scale food dataset Food-101 from foodspotting.com for  food image recognition. \cite{Sajadmanesh-KC-arXiv2016} used  large-scale recipes from Yummly with 157,013  recipes from over 200 types of cuisines to analyze worldwide culinary habits. \cite{Min-YAWYE-TMM2018} further combined  food images with rich recipe attribute information from Yummly to analyze and compare culinary cultures in different geographical regions.  In addition, there are rich social information provided by  some recipe websites, such as Allrecipes and Epicurious, e.g., ratings and comments, which can be helpful for many tasks, e.g., recipe recommendation \cite{Teng-RRUIN-WSC2012} and recipe rating prediction \cite{Yu2013Do}.

Besides recipe-sharing websites, the social media, such as Twitter, Facebook, Foursquare, Instagram and Youtube also provide large-scale food data. For example, \cite{Culotta2014Estimating} examined whether linguistic patterns in Twitter correlate with health-related statistics. \cite{Abbar2015You} combined the demographic information and food names from Twitter  to model the correlation between calorie value and diabetes. In addition to  textual data, recent studies \cite{Mejova2016Fetishizing,Ofli-Saki-WWW2017} have used large-scale food images from social media for the study of  food perception and eating behaviors.

Collecting  food data from  IoT devices is also a common way.  With the popularity of cameras embedded in smartphones and various wearable devices \cite{Vu2017Wearable}, researchers begin capturing food images in restaurants or canteens for visual food understanding \cite{Farinella2015A,Ciocca2016Food,Damen2018EPICKITCHENS}. For example, \cite{Ciocca2016Food} collected  food images in a real canteen  using the smart phone. Besides  food images, \cite{Damen2018EPICKITCHENS} used the head-mounted GoPro to record the cooking videos.

There are also some work, which collected  food data via  crowd-sourcing platforms, such as the Amazon Mechanical Turk (AMT). For example, \cite{Noronha2011Platemate} introduced a system to crowdsource nutritional analysis from photographs using AMT. \cite{kawano2014automatic} proposed a framework for automatic expansion based on some food image seeds  via the AMT and the transfer learning. \cite{Lee-MRMCFD-KDD2017} built a large dataset of less structured food items via crowd-sourcing and  conducted the match between the formally structured restaurant menu item and this dataset.

In summary, food-related data from different sources is divided into the following several types:
\begin{myitemize}
\item Recipes: Recipes contain a set of ingredients and sequential instructions. In the earlier research, recipes are collected from cookbooks and manually typed into computers. Currently, recipes can be collected from many cooking websites, such as epicurious.com and Allrecipes.com. As a result, their numbers have grown exponentially. Recipes can be embedded in the latent space for recipe analysis and other tasks. For example, \cite{Teng-RRUIN-WSC2012} proposed an ingredient network for recipe recommendation. \cite{Kim2016Tell} used the recipes to examine correlations between individual food ingredients in recipes.
\item Dish images: Dish images are the most common multimedia data due to their visual information and semantic content. They contain food images  with categories, and we can  extract meaningful concepts and information via existing deep learning methods for various  food tasks. Most tasks conduct the visual analysis for food images with the single item. There are also some food image datasets such as  UEC Food256~\cite{kawano2014automatic} and UNIMIB2016~\cite{Ciocca2016Food}  with multiple food-items. Figure \ref{dataset_samples}~(b)  shows some examples.
\item Cooking videos: Nowadays, there are many cooking videos, which can guide person how to cook. They contain human cooking activities and cooking procedure information.  Many researchers can use such data for  human cooking activity recognition and other tasks~\cite{Damen2018EPICKITCHENS}.
\item Food attributes: Food contains rich attributes, such as  flavors, cuisine, taste, smell, cooking and cutting attributes. We can adopt  rich food attributes  to improve food recognition and other tasks. For example, \cite{Jing-CMR-MM2017} conducted food recognition and recipe retrieval via  predicting rich food attributes, such as ingredient and cooking attributes.
\item Food log: Food log records rich information, including food images, text and other calorie information. With the rapid growth of mobile technologies and applications, we can use the FoodLog App  to  keep the healthy diet. Some works such as \cite{Kitamura2008Food} introduced a food-logging system for the analysis of  food balance.
\item Restaurant-relevant food information: Nowadays, more and more works use restaurant-specific information, including the menu and GPS information for restaurant-specific food recognition. Such data type includes the dish list from the restaurant. For example, \cite{XuRuihan-GMDR-TMM2015}  utilized the menu data type for discriminative classification in geolocalized settings.
\item Healthiness: More and more people pay  attention to the health because of the improved living standard. The healthiness  contains rich information, such as the  calorie and nutrition. An excessive unhealthy lifestyle and bad dietary habits can trigger overweight, obesity and other diseases. Researcher can use the healthiness of food to keep the healthy diet. For example, \cite{Okamoto-ACES-MM2016} proposed a calorie estimation system  for automatic food calorie estimation from the food image.
\item Other food data: Other food data includes the data from cooking books or the government, questionnaire, odor threshold database\footnote{http://www.thresholdcompilation.com/}, food product codes and so on.
\end{myitemize}

\textbf{Existing Benchmark Food Datasets.} Many benchmark and popular food datasets are also constructed and released. Table \ref{datasets_1} and Table \ref{datasets_2} list main food-related databases in more details, where the number in () denotes the number of categories. We also give the links for datasets if they are available. From Table \ref{datasets_1} and Table \ref{datasets_2},  we can see that
\begin{itemize}
\item The benchmark datasets for food image recognition are released frequently. Earlier, researchers focus on  the food dataset with few cuisines and small-scale. For example, UEC Food100 \cite{Matsuda2012Multiple} consists of 14,361  Japanese food images. Benefiting from the fast development of  social media and mobile devices, we can easily obtain larger amounts of food images. For example,  \cite{Rich-TBUASF-ICDH2016} released a dataset with 808,964 images from Instagram. In addition, ETHZ Food-101~\cite{Bossard-Food101-ECCV2014} has been a benchmark food dataset for the food recognition task.
\item There are  some  restaurant-oriented datasets, such as Dishes~\cite{XuRuihan-GMDR-TMM2015} and Menu-Match~\cite{Beijbom-MeMa-WACV2015}. Such datasets generally contain the location information, such as GPS or  restaurant information.
\item Compared with  food images, recipes contain richer attribute and metadata information. To the best of our knowledge, Recipe1M \cite{Salvador-LCME-CVPR2017}  is the largest released recipe dataset, which contains  1M cooking recipes and 800K food images.  Recently, \cite{Semih-RecipeQA-EMNLP2018} released a recipe dataset RecipeQA, which includes additional 36K questions to support question answering compared with other recipe datasets. Some datasets with the cooking videos are also released for human-activity recognition and prediction, such as recently released EPIC-KITCHENS~\cite{Damen2018EPICKITCHENS}.
\end{itemize}

\begin{table*}[htbp]
\caption{Food-related Datasets.}
	\begin{center}
        \scalebox{0.80}[0.80]{
		\begin{tabular}{|c|c|c|c|c|c|}
			\hline
			Reference&Dataset Name&Data Type& Num.& Sources& Tasks  \\
            \hline
			\cite{Chen2009PFID}&PFID& Images with categories & 4,545 (101) & Camera & Recognition\\
            \cite{Joutou2010A}&Food50& Images with categories & 5,000 (50) & Web & Recognition \\
			\cite{Hoashi2010Image}&Food85& Images with categories & 8,500 (85)&  Web& Recognition \\
            \cite{Chen2012Automatic} &-& Images with categories & 5,000 (50)&  Web+Camera& Quantity Estimation\\
            \cite{Matsuda2012Multiple}&UEC Food100$^{1}$ &Images with categories& 14,361(100) & Web+Manual &Recognition \\
            \cite{Anthimopoulos2014A}&Diabetes&Images with categories& 4,868(11) & Web &Recognition\\
            \cite{kawano2014automatic} &UEC Food256$^{2}$ &  Images with categories & 25,088(256)& Crowd-sourcing & Recognition\\
			\cite{Bossard-Food101-ECCV2014} &ETHZ Food-101$^{3}$&Images with categories & 10,1000(101) & foodspotting.com  & Recognition\\
            \cite{Wang-RRLMFD-ICME2015} &UPMC Food-101$^{4}$& \tabincell{c}{Images and text \\with categories}  & 90,840(101)& Google Image search & Recognition\\
			\cite{Farinella2015A}&UNICT-FD889$^{5}$&Images with categories& 3,583(889) & Smartphone & Retrieval\\
			\cite{Pouladzadeh2015FooDD}&FooDD$^{6}$&Images with categories &3,000(23) & Camera &  Detection\\
            \cite{Christodoulidis-FRDA-ICIAP2015}&-&Images with categories&(572)& Manual&Recognition\\
            \cite{Meyers-Im2Calories-ICCV2015}&Food201-Segmented& Images with categories&12,625(201)&Manual&Segmentation\\
            \cite{Bettadapura-LCSAFR-WACV2015}&-& \tabincell{c}{Images with \\categories and location}& 3,750(75)&  Web&Recognition \\
			\cite{XuRuihan-GMDR-TMM2015}&Dishes$^{7}$&\tabincell{c}{ Images with\\ categories and location} & 117,504(3,832) &Dianping.com &Recognition\\
            \cite{Beijbom-MeMa-WACV2015}&Menu-Match$^{8}$& \tabincell{c}{Images with categories}& 646(41) & Social media&Food Logging\\
            \cite{Ciocca2015Food}&UNIMIB2015$^{9}$&Images with categories & 2000(15)&Smart phone&Recognition\\
            \cite{Ciocca2016Food}&UNIMIB2016$^{9}$&Images with categories & 1,027(73)&Smart phone &Recognition\\
            \cite{Zhou2016}&Food-975&Images with categories& 37,785(975) &   Camera\&yelp &  Recognition\\
            \cite{Merler2016} &Food500& Images with categories& 148,408 (508) & Web\&Social media &Recognition\\
            \cite{Rich-TBUASF-ICDH2016}&Instagram800K$^{10}$& Images with tags & 808,964(43)& Instagram &  Recognition\\
            \cite{Singla-FnFC-MM2016}&Food11 & \tabincell{c}{Images with categories} & 5,000 (50) & Social media & Recognition \\
            \cite{Farinella2016Retrieval}&UNICT-FD1200$^{11}$& Images with categories &4,754(1,200) & Mobile camera &\tabincell{c}{Recognition\\ and Retrieval}\\
            \cite{Ofli-Saki-WWW2017}&-& Images with tags & 1.9M& Instagram &  Food Perception\\
            \cite{liang2017computer}&ECUSTFD$^{12}$ & \tabincell{c}{Images with \\ rich annotation} & 2978(19) &Smart phone &\tabincell{c}{Calorie\\ Estimation}\\
            \cite{ciocca2017learning}&Food524DB$^{13}$& Images with categories &247,636(524) & Existing datasets &Recognition\\
            \cite{Muresan-FR-Acta2018}&Fruits 360 dataset$^{14}$& Fruit images with categories &71,125(103)& Camera&Recognition\\
            \cite{Hou-VegFru-ICCV2017}&VegFru$^{xx}$& Fruit and vegetable images with categories &160,731(292)& Search Engine&Recognition\\
            \cite{Georg-PDSM-ICIAP2017}&FruitVeg-81$^{xxx}$& Fruit and vegetable images with categories &15,630(81)& mobile phone&Recognition\\
            \cite{Chen2017ChineseFoodNet}&ChineseFoodNet$^{15}$&Images with categories&192,000(208) &Web &-\\
            \cite{thanh2017healthy}&Instagram 1.7M&Images with comments & 1.7M & Instagram&\tabincell{c}{Consumption Patterns \\Analysis}\\
            \cite{Harashima2017Cookpad}&Cookpad$^{16}$&Images and recipes &\tabincell{c}{4,748,044}&Cookpad&-\\
            \cite{Qiang-BTBUFood60-bigcomp2019}&BTBUFood-60&Images with bounding-box annotation &\tabincell{c}{52,495}&Baidu\&Google&Food Object Detection\\
            \hline
         \end{tabular}
         }
    \end{center}
    \begin{tablenotes}
    \footnotesize
       \item[1] $^{1}$\url{http://foodcam.mobi/dataset100.html/}.
       \item[2] $^{2}$\url{http://foodcam.mobi/dataset256.html/}.
       \item[3] $^{3}$\url{http://www.vision.ee.ethz.ch/datasets_extra/food-101/}.
       \item[4] $^{4}$\url{http://visiir.lip6.fr/}.
       \item[5] $^{5}$\url{http://iplab.dmi.unict.it/UNICT-FD889/}.
       \item[6] $^{6}$\url{http://www.site.uottawa.ca/~shervin/food/}.
       \item[7] $^{7}$\url{http://isia.ict.ac.cn/dataset/Geolocation-food/}.
       \item[8] $^{8}$\url{http://neelj.com/projects/menumatch/}.
       \item[9] $^{9}$\url{http://www.ivl.disco.unimib.it/activities/food-recognition/}.
       \item[10] $^{10}$\url{http://www.eecs.qmul.ac.uk/~tmh/downloads.html}.
       \item[11] $^{11}$\url{http://www.iplab.dmi.unict.it/UNICT-FD1200/}
       \item[12] $^{12}$\url{https://github.com/Liang-yc/ECUSTFD-resized-}.
       \item[13] $^{13}$\url{http://www.ivl.disco.unimib.it/activities/food524db/}.
       \item[14] $^{13}$\url{https://www.kaggle.com/moltean/fruits}.
       \item[15] $^{14}$\url{https://sites.google.com/view/chinesefoodnet/}
       \item[16] $^{15}$\url{https://www.nii.ac.jp/dsc/idr/cookpad/cookpad.html}
       \item[xx] $^{xx}$\url{https://github.com/ustc-vim/vegfru}
       \item[xxx] $^{xxx}$\url{https://www.tugraz.at/institute/icg/research/team-bischof/lrs/downloads/fruitveg81/}
    \end{tablenotes}
\label{datasets_1}
\end{table*}

\begin{table*}[htbp]
\caption{Continued}
	\begin{center}
        \scalebox{0.80}[0.8]{
		\begin{tabular}{|c|c|c|c|c|c|}
			\hline
			Reference&Dataset Name &Data Type& Num.& Sources& Tasks  \\
            \hline
            \cite{Rohrbach-FGAD-CVPR2012}&MPII Cooking 2$^{16}$&Cooking videos&273 & Cameras &\tabincell{c}{Cooking Activity \\Recognition}\\
            \cite{Stein-CEA-CPUC2013}&50 Salads$^{17}$& Cooking videos & 50 & Cameras &\tabincell{c}{Cooking Activity \\Recognition}\\
            \cite{Kuehne-LA-CVPR2014}&Breakfast$^{18}$& Cooking videos & 433 & Cameras &\tabincell{c}{Cooking Activity \\Recognition}\\
            \cite{Damen2018EPICKITCHENS}&EPIC-KITCHENS$^{19}$& Cooking videos& 432  & \tabincell{c}{Head-mounted\\ GoPro}&\tabincell{c}{Cooking Activity \\Recognition}\\
			\hline
            \cite{Kinouchi2008The}&-&Recipes&\tabincell{c}{7,702}&-&Culinary Evolution\\
            \cite{Ahn-FNFP-SciRe2011}&Recipes56K$^{20}$&Recipes & 56,498 &Recipe websites&\tabincell{c}{Ingredient Pattern \\Discovery}\\
            \cite{Teng-RRUIN-WSC2012}&-&Recipes&\tabincell{c}{46,337}&allrecipes.com&Recipe Recommendation\\
            \cite{Kim2016Tell}&-&Recipes&5,\tabincell{c}{917}&Recipesource.com&\tabincell{c}{ Recipe Analysis}\\
            \cite{Chen-DIRCRR-MM2016}&Vireo Food-172$^{21}$&\tabincell{c}{Recipes with \\images and ingredients}&110,241(172)&Web and manual&Recipe Retrieval\\
            \cite{Sajadmanesh-KC-arXiv2016}&Recipes157K& Recipes with metadata & 157K& Yummly &\tabincell{c}{Cross-region Food \\Analysis}\\
			\cite{Chen-CMRR-MMM2017}&Go cooking& Recipes\&Images & 61,139& xiachufang.com &\tabincell{c}{Cross-modal \\Recipe Retrieval}\\
			\cite{Salvador-LCME-CVPR2017}&Recipe1M$^{22}$& Recipes\&Images &1M & Cooking websites & \tabincell{c}{Cross-modal \\Recipe Retrieval}\\
            \cite{WeiqingMin-BSC-TMM2017}&Yummly-28K$^{23}$& Recipes\&Images & 28K& Yummly &\tabincell{c}{Cross-modal\\ Retrieval}\\
            \cite{Min-YAWYE-TMM2018}&Yummly-66K$^{24}$& Recipes\&Images & 66K& Yummly &\tabincell{c}{Cross-region Food \\Analysis}\\
            \cite{Markus-Recipe-ICWSM2018}&Recipes242K$^{25}$&Recipes&242,113 & Crowdsourcing & \tabincell{c}{Recipe Healthiness\\ Estimation}\\
            \cite{Semih-RecipeQA-EMNLP2018}&RecipeQA$^{26}$&Recipes&~20K(22)&instructables.com& \tabincell{c}{Recipe\\Question Answering}\\
            \hline
         \end{tabular}
         }
    \end{center}
    \begin{tablenotes}
    \footnotesize
    \item[16]$^{16}$\url{https://www.mpi-inf.mpg.de/departments/computer-vision-and-multimodal-computing/research/human-activity-recognition/mpii-cooking-2-dataset/}
    \item[17]$^{17}$\url{http://cvip.computing.dundee.ac.uk/datasets/foodpreparation/50salads/}.
    \item[18]$^{18}$\url{http://serre-lab.clps.brown.edu/resource/breakfast-actions-dataset/#Downloads}.
    \item[19]$^{19}$\url{https://epic-kitchens.github.io/2018}.
    \item[20]$^{20}$\url{http://www.yongyeol.com/2011/12/15/paper-flavor-network.html}
    \item[21]$^{21}$\url{http://vireo.cs.cityu.edu.hk/VireoFood172/}.
    \item[22]$^{22}$\url{http://im2recipe.csail.mit.edu/}.
    \item[23]$^{23}$\url{http://isia.ict.ac.cn/dataset/}.
    \item[24]$^{24}$\url{http://isia.ict.ac.cn/dataset/Yummly-66K.html}.
    \item[25]$^{25}$\url{https://github.com/rokickim/nutrition-prediction-dataset/blob/master/}.
    \item[26]$^{26}$\url{https://hucvl.github.io/recipeqa}
    \end{tablenotes}
\label{datasets_2}
\end{table*}

These increasing amount of food-related data presents researchers with more opportunities for  food analysis. Such analysis can be conducted not only on  these data sets individually, but  also multiple datasets jointly. For example, we can analyze the correlation between chemical data and recipes \cite{Ahn-FNFP-SciRe2011} or social media images and obesity \cite{Mejova2016Fetishizing}. These connections with different kinds of food data can provide us with a new perspective on the study of food from different angles, such as the culinary habits and human behavior.

\section{Applications in Food Computing} \label{section_AFC}
Before introducing  core   tasks in food computing, we first list a number of applications and summarize them from the following main four aspects: health, agriculture, culture and food science.

\subsection{Health}
What kind of food or how much we eat is closely related to our health. For example, if we eat too much, we can  at risk for developing many diseases, such as  diabetes and heart disease. Therefore, food-relevant study will benefit various health-oriented applications. We next introduce four representative food-oriented health applications, including (1) food perception for health, (2) food recognition for diet  management, (3) health-aware food recommendation and  (4) food-health analysis from social media.

\subsubsection{Food Perception for Health.} How we choose  food  is very relevant to how we perceive food, e.g., whether it is fresh or tasty. Due to the global overweight and obesity epidemic, an increasing number of researchers  studied how we perceive food, both before and during its consumption. For example, \cite{S2003Effect} studied the relation between  the positive sensory food perception and intake. \cite{Mccrickerd2016Sensory} focused on how multimodal cues (visual and odor) affected the food identification and the guidance of food choice. \cite{Ofli-Saki-WWW2017} used the image recognition method to compared the difference between how a human labels food images and how a machine labels them, and then discovered some facts, such as the positive correlation between the food choice and regions with better health outcomes.

\subsubsection{Dietary  Management for Health.} Dietary assessment or food diary \cite{Lydia-TBYE-INCS2008,Cordeiro2015Barriers,Cordeiro-RMFJ-HFCS2015,Cordeiro2015Barriers,Achananuparp2018Does} provides valuable insights  for preventing many diseases. Traditional methods  mostly rely on questionnaires or self-reporting \cite{Thompson2008Dietary}. These methods have many problems due to underreporting and miscalculation of food consumption.  With  the development of computer vision methods, more approaches resort to vision-based methods  for diet management.

To the best of our knowledge, the first method for food intake analysis of the digital photography is developed by \cite{Williamson2003Comparison,Williamson2004Digital}, which  measured the food intake in the cafeteria settings. The system used a digital video camera to capture a photograph of a participant's food selection before they eat, and plate waste after they finish eating. These photographs are then analyzed by registered dietitians. These portion size estimates are entered into a custom built computer application that automatically calculates the grams and kilocalories of food selection based on the  USDA database. \cite{Martin2009Quantification} further developed a  remote food photography method in the natural environment. In contrast, \cite{Noronha2011Platemate} introduced a system PlateMate for crowdsourcing nutritional analysis from photographs of meals.

These methods are semi-automatic and involve the participant of registered dietitians. To make the system full-automatic, \cite{Zhu2008Technology,Zhu2010The} proposed  a technology assisted dietary assessment system, where  images obtained before and after foods are eaten, are used to estimate the type and  amount of food consumed. Similar methods including sing-view reconstruction and multi-view reconstruction for food volume estimation \cite{Xu2013MODEL,Almaghrabi2012A,Chen2013Model,Shevchik2013Food,Pouladzadeh2014Measuring,Dehais2017Two} are proposed. Such methods generally need  3D reconstruction from food images.

\begin{figure*}
\centering
\includegraphics[width=0.75\textwidth]{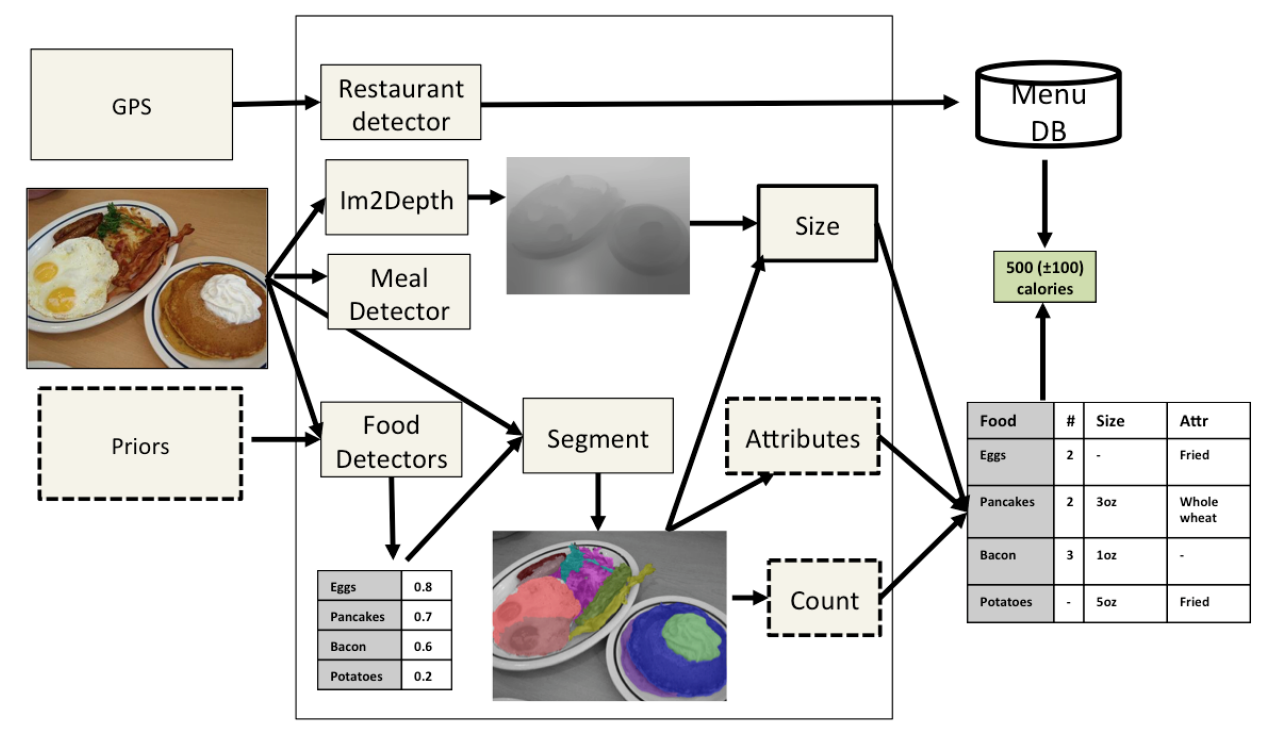}
\caption{The framework of Im2Calories~\cite{Meyers-Im2Calories-ICCV2015}}
\label{Im2Calories}
\end{figure*}

Recently, a lot of work focused on  calorie estimation from one image. For example, \cite{Pouladzadeh2014Mobile} provided accurate food recognition and calorie measurement by periodically training the system with more food images. \cite{Meyers-Im2Calories-ICCV2015} proposed an Im2Calories system (Figure \ref{Im2Calories}), which used  a segmentation based approach to localize the meal region of the food  photo, and  then applied the multilabel classifiers to label these segmented regions.  Once the system  segmented the foods, it can  estimate their volume. \cite{Okamoto-ACES-MM2016} proposed an image-based calorie estimation system, which estimates the food calories automatically by simply taking a meal photo from the top with a pre-registered reference object. \cite{akpa2017smartphone} automatically measured food  calories from food photo using ordinary chopsticks as a measurement reference. \cite{Ege2017Image} estimated the food calorie from a food photo via simultaneous learning of food calories, categories, ingredients and cooking directions using multi-task Convolutional Neural Networks (CNNs)\cite{Krizhevsky2012ImageNet}. Recently, \cite{Fang2018Single} presented a food portion estimation method to estimate the energy from food images using  generative adversarial networks.

Mobile phones are becoming a popular and powerful platform, more works conduct food calorie estimation on  mobile devices. Both \cite{Parisa-YAWYE-Instrum2016} and \cite{BVR-AFM-JHMI2017} gave a survey to present a variety of methods on automated food monitoring and dietary management. Please refer to them in more details. In addition, there are also some works on other devices. For example, \cite{Chen2012Automatic} estimated food categories and volumes by the depth cameras such as Kinect. Several wearable devices, such as glasses with load cells \cite{Chung2017A} or connected to sensors on temporalis muscle and accelerometer \cite{Farooq2016A}, have been explored to detect food intake events automatically. The collected information about eating episodes, pertinent to users' diet habit pattern, can serve as a starting point for food consumption analysis and diet interventions, e.g., providing user recommendations for healthier food and eating habit \cite{Faiz2015An}.  Please refer to \cite{Vu2017Wearable} in more details on wearable food intake monitoring.  Some work \cite{Waki-MFRT-DST2015}, \cite{Kitamura2008Food}\cite{Miyazaki2011Image} proposed a food-logging system that can distinguish food images from other images, and analyze the food balance. \cite{Aizawa-FoodLog-IEEEMM2015} created a multimedia food-recording tool FoodLog, which offers a novel method for recording our daily food intake  for healthcare purposes. \cite{Beijbom-MeMa-WACV2015} focused on the restaurant scenario and present an automated computer vision system for logging food and calorie intake using images. \cite{Goyal2017I} proposed a project to provide a convenient, visual and reliable way to help users learn from their eating habits and guide them towards better food choice.

Besides food recognition for the dietary  management, some researchers designed different sensors to looking to track their diets and count their calories\cite{Strickland-Sensors-IEEESp2018}. For example, some researchers have designed the sensor, which can stick to the uneven surface of a tooth to monitor the wearer's glucose, salt, and alcohol intake. \cite{Min-AMA-MSys2018} explored an audio-kinetic model of well-formed multi-sensory earable devices for dietary monitoring. In addition, there are also some works, which applied constraint reasoning  to dynamically adapt dietary recommendations for compensating diet transgressions \cite{Luca-TDM-AnselmaM15,Anselma-AIF-JBI2017,Anselma-ARE-MADiMa2018}.

\subsubsection{Health-aware Food Recommendation.} When motivating research on food recommender systems, health problems and improving eating habits are usually mentioned. For example,  \cite{Ge-UTLF-ICDH2015}  took calorie counts into consideration in the recommendation algorithm  based on their proposed ``calorie balance function'' that can account for the difference between  calories the user needs and  ones in a recipe. \cite{Harvey2015Automated} realized the trade-off for most users between recommending the user what she wants and what is nutritionally appropriate. \cite{Harvey2017Exploiting} employed a post-filtering approach to incorporate  nutritional aspects. There is still large space for improving health-aware food recommendation to provide better health service by taking  more factors, e.g., the user intake, the estimation of portion size, and other personal factors into account.

\subsubsection{Food-Health Analysis from Social Media.} We're in an era of social media. As food is  indispensable to our life, a great deal of online content is relevant to food. We upload food photos, find recipes and talk about them. Therefore, a great amount of information about our culinary habits and behavior is  being recorded via the social media. Recent studies have shown that we can  use social media to get aggregating statistics about the health of people for public health monitoring. For example, \cite{Culotta2014Estimating} presented a large-scale study of 27 health-related statistics, such as the health insurance coverage and obesity. \cite{Fried2014Analyzing}  collected a large corpus of food-related tweets in Twitter and used all these tweets to predict latent population characteristics such as  overweight and diabetes rates. \cite{Abbar2015You} used  daily food-related tweets  from Twitter to predict the national obesity and diabetes statistics. \cite{Mejova-FoodPorn-ICDH2015} used the Foursquare and Instagram images to study food consumption patterns in the US, and find the correlation between obesity and fast food restaurants.  \cite{Mejova2016Fetishizing} connected  food perception and food images for public health analysis via social media.

\subsection{Culture}
Food is fundamental to the culture, with food practices reflecting our nationalities and other aspects \cite{Khanna-FC-EFN2009,Harris-G2E-AA1985,Bell1997Consuming,Giampiccoli-TFC-CA2012}. An understanding of food culture is indispensable in  human communication. This is true not only for professionals in many fields such as  public health and  commercial food services, but is clearly recognized  in the global marketplace. Food has also come to be recognized as part of the local culture which tourists consume, as an element of regional tourism promotion and a potential component of local agricultural and economic development \cite{Hall2003Wine}. In addition, exploring the food culture can help  develop personalized food recommendation  considering the  aspect of  food culture  from different urban areas.

For these reasons, more and more work focused on the study of  culinary cultures. For example, \cite{Zhu2013Geography} harnessed  online  recipes from various Chinese regional cuisines and investigated the similarity of regional cuisines in terms of geography and climate. \cite{Ahn-FNFP-SciRe2011} introduced a flavor network from  recipes to identify  significant ingredient patterns that indicate the way humans choose  paired ingredients in their food. These patterns  vary  from geographic region to geographic region. For example, these ingredients with shared flavor compounds tend to be combined for North American dishes. \cite{Simas2017Food} analyzed and discussed a possible new principle behind traditional cuisine: the Food-bridging hypothesis and its comparison with the food-pairing hypothesis using the same dataset and graphical models from \cite{Ahn-FNFP-SciRe2011}. \cite{Strohmaier2015Mining}  proposed an approach to mine cultural relations between different language communities through their description of interest in their own and other communities' food culture. \cite{Herder2016Plate} provided  large-scale empirical evidence on gender differences in cooking behavior. \cite{Kim2016Tell}developed  a food analyzer, which used the data from recipes to examine correlations between individual food ingredients in recipes. This paper found that  meaningful correlations characterizing the food culture of each area can be explained by these authentic ingredients in recipes. \cite{Sajadmanesh-KC-arXiv2016} used  large-scale recipes from Yummly to analyze and compare  worldwide cuisines and culinary habits.

\begin{figure*}
\centering
 \includegraphics[width=0.90\textwidth]{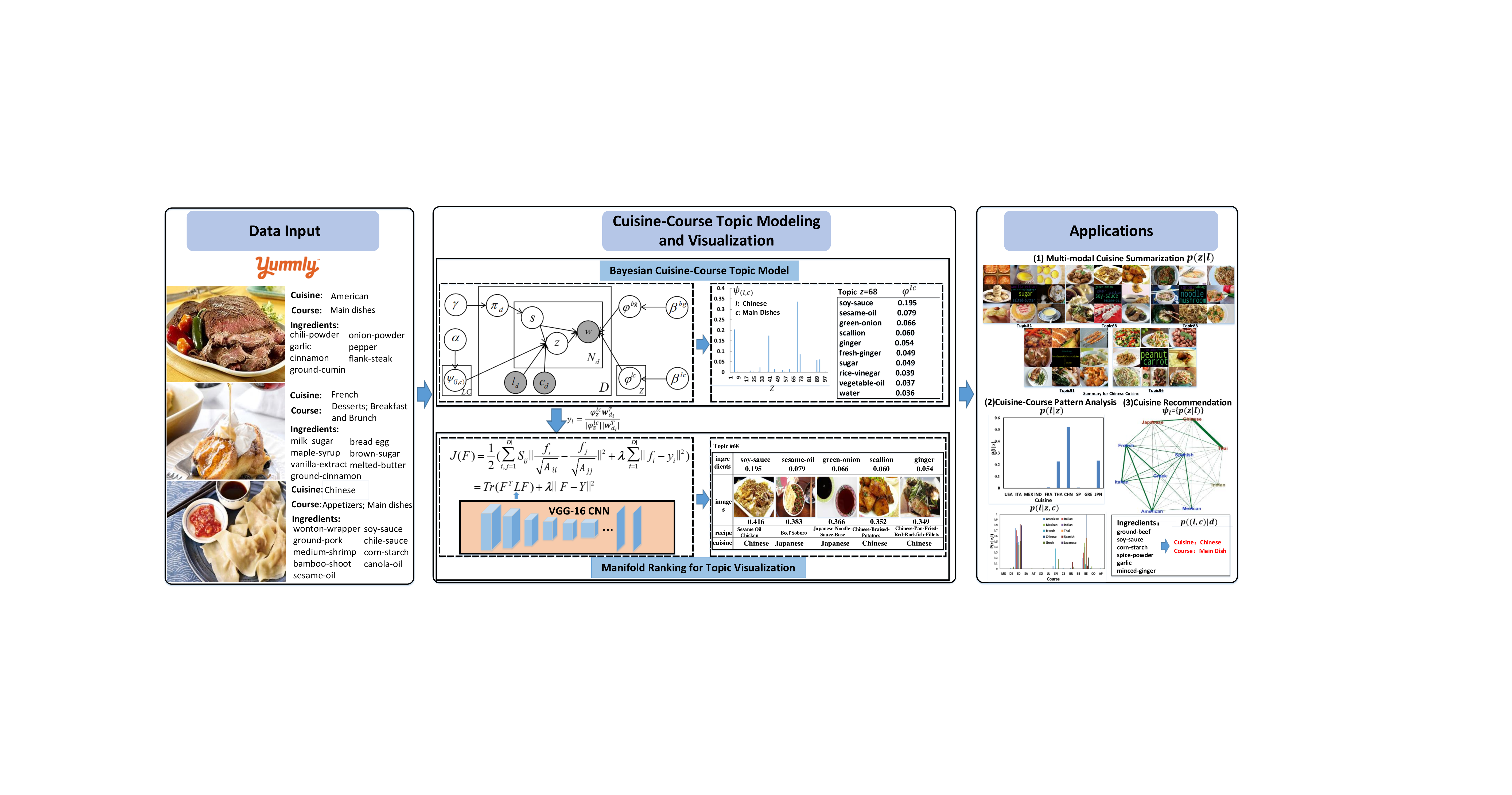}
\caption{The proposed  culinary culture analysis  framework \cite{Min-YAWYE-TMM2018}.}
\label{BC2TMV}       
\end{figure*}
As  representative work, \cite{Min-YAWYE-TMM2018} recently combined  food images with rich recipe attribute information from Yummly for the comparison of culinary cultures. Fig. \ref{BC2TMV}  shows the proposed  analysis  framework, where  the input is  recipe information, including   food images, ingredients and various attributes (e.g., the cuisine and course) from Yummly. They first proposed a Bayesian  topic model to discover  cuisine-course specific topics. Based on  the learned   distribution, they then retrieved relevant food images for topic visualization. Finally, the authors exploited the topic modeling and visualization for cross-region food analysis at both macro-level and micro-level.

Besides recipes, social media based food culture analysis has been conducted for both food culture understanding and popular food culture prediction \cite{Abbar2015You,Ofli-Saki-WWW2017}.  \cite{Yanai2009Detecting} generated representative geotagged food photographs from typical regions in the world to find the cultural difference. \cite{Silva-YAWYE-Arxiv2014} used 5 million Foursquare check-ins to find  strong temporal and spatial correlation between individuals' cultural preferences and their eating and drinking habits. \cite{Abbar2015You} used tweets from Twitter to study  dietary habits.  \cite{Ofli-Saki-WWW2017} found that the conscious food choices are probably  associated with regions of better health outcomes. The prosperity of  social media provides opportunities to obtain  detailed records of individual food consumption from millions of people, which will continue revolutionizing our understanding of food choice and culinary culture, and their impact on health and well-being.

\subsection{Agriculture}
Food computing can also be used in the agriculture or food products. Food image analysis  has many potential applications for automated agricultural and food safety tasks \cite{Chen2002Machine,Senthilnath2016Detection,Xiang2014Recognition}. For example, \cite{Jim1999Automatic} proposed a  recognition system, which uses a laser range-finder model and a dual color/shape analysis algorithm to locate the fruit. Recently, artificial vision systems have become  powerful tools for automatic recognition of fruits and vegetables. For example, \cite{Hern-SOC-ICTI2017} presented the image capture, cropping and process for fruit recognition. \cite{Chen2017Counting} introduced a deep learning method to extract visual features for counting fruits. \cite{Hossain-AFC-TII2019} introduced a deep learning method to extract visual features for automatic  fruit recognition for industry applications. \cite{Lu-Innovative-AS2017} provided a brief overview of hyperspectral imaging configurations and common sensing modes for food quality and safety evaluation. For natural food product classification, \cite{Patel1998Color} proposed a  neural network to accurately distinguish between grade A eggs and blood spot eggs. \cite{Pabico2015Neural} used an artificial neural network to automate the classification of tomato ripeness and acceptability of eggs. \cite{Chatnuntawech-arXiv2018} developed a non-destructive rice variety classification system, which used a hyperspectral imaging system to  acquire complementary spatial and spectral information of rice seeds, and then used  a deep CNN to extract the features from  spatio-spectral data  to determine the rice varieties. Recently, there are also some work, which uses the visual information to evaluate the food quality\footnote{https://hackernoon.com/your-pizza-is-good-how-to-teach-ai-to-evaluate-food-quality-d835a8c12e86}.

It is worth noting that agriculture-oriented food recognition is more similar to object  recognition in the computer vision, such as fruit recognition \cite{Hern-SOC-ICTI2017} and egg classification \cite{Patel1998Color}. However, it is quite different from dish or ingredient recognition. In contrast to object-like recognition, food typically does not exhibit any distinctive semantic parts: while we can decompose one object such as  bird into some fixed semantic parts, such as head and breast, we cannot find similar semantic patterns from one  dish. As a result, we should design new recognition methods or paradigms for dish or ingredient recognition.

\subsection{Food Science}
According to Wikipedia, food science is defined  as the application of basic sciences and engineering to study the physical, chemical and biochemical nature of foods and principles of food processing\footnote{https://en.wikipedia.org/wiki/Food\_science}. Food computing provides new methods and technologies for these sub-areas. For example, sensory analysis is to study how human senses perceive food. Food perception  uses the Magnetic Resonance Imaging (MRI)  to measure brain activity based perception, and thus is often conducted in the lab \cite{Killgore2005Body}.
In contrast, \cite{Ofli-Saki-WWW2017} considered this problem as food image recognition  from Instagram and showed the perception gap between how a machine  labels an image and  how a human  does. In addition, food perception should be multi-modal and it includes visual and auditory cues, tastes, smells and tactile sensations. Therefore, multi-modal integration is needed. Existing studies \cite{Verhagen2006The} focused on this topic from the neuroscience. However, we can resort to deep learning based multimodal learning methods \cite{Srivastava2012Multimodal} in computer science to better tackle this problem. Another example is the quality control. Some works \cite{Pabico2015Neural} used the neural network to automate the classification of tomato ripeness and acceptability of eggs.

\section{Tasks  in  Food Computing}\label{section_FCT}
In this section, we introduce main tasks of food computing, including perception, recognition, retrieval, recommendation, prediction and monitoring.

\subsection{Perception}
One important aspect determining our food choice and how much we eat/drink is how we perceive food from its certain characteristics, such as whether it is sweet or tasty. Therefore, the study on food perception plays an important part in our health. In addition, such  study  will have a number of important potentials for  food and beverage industries,  for example, a better understanding of the process used by people to assess the acceptability and flavor of new food products.

Many studies on food perception  are conducted at the level of brain activity typically in labs. For example, \cite{Killgore2003Cortical} found that healthy, normal weight participants are presented with color photographs of food or food-related stimuli differing in caloric value (e.g., high-calorie or low-calorie food) while they underwent blood oxygen-level-dependent  functional MRI. \cite{S2003Effect} gave a review on the connection between  sensory food perception  and human eating behavior.  \cite{Rosenbaum2008Leptin} demonstrated that after undergoing substantial weight loss, there are some changes in brain activity because of the elicitation from food-related visual cues for obese subjects. \cite{Nenad2016The}  found that both lean and overweight subjects showed similar patterns of  neural responses to some attributes of food, such as health and taste.

There are also some work which is more directly related to visual perception of food. For example, \cite{Spence2010Does} studied the influence of  food color on perceiving the taste and flavor.  \cite{Ofli-Saki-WWW2017} used the  image recognition method to study the relation between how food is perceived and what it actually, namely the food perception gap. The food perception actually involves multi-modalities, including  not only visual and auditory cues, but also  tastes, smells and tactile sensations. When we are chewing food, we can perceive the taste, flavor or texture, which will  facilitate our appreciation of food. The senses of taste and smell play a great role in choosing  food. Visual information of a food product is essential in the choice and acceptance of this product, while auditory information obtained during the chewing of food products will help us judge whether a product is fresh or not. Food perception does not just depend on one  sense, but should be the result from multisensory integration on various types of signals. Therefore, human  food perception is multimodal. For example, \cite{Mccrickerd2016Sensory} studied the role of  multimodal cues including  both visual and odor ones in recognizing food and selecting food. Particularly, they described the affect of  the size of a plate or the amount of food served on the food intake.  \cite{Verhagen2006The} reviewed existing works on  multimodal food perception and its neurocognitive bases. In addition,  some work such as \cite{MOURITSEN201716} used the multivariate data analysis method to predict reasonably  sensory properties from  chemical characterization of sauces.

In summary, food perception has received the rapid growth of research interest especially in the neuroscience, cognition and health-related fields. Most  methods are  domain-specific. Advanced computer vision and machine learning methods in computer science have not been fully exploited for food perception. Note that recently, some work \cite{Ofli-Saki-WWW2017} is beginning utilizing  big data from  social media and computer vision from AI for the study of food perception. With the fast development of AI and its combination with neuroscience, we believe more and more methods from computer science will be applied to food perception. For example, one important problem of multimodal food perception is that how multimodal features of food are integrated effectively. A feasible method is to employ existing deep learning networks, such as \cite{Srivastava2012Multimodal,Ngiam2009Multimodal} for effective fusion on  heterogeneous signals.

\subsection{Recognition}
Diet-related chronic diseases like obesity, diabetes and cardiovascular disease have become a major health concern. Diet management is a key factor to prevent and treat these diseases. Traditional food diary methods  manually recorded types and portion of consumed food, and thus the accuracy is hard to guarantee.  The widespread use of smartphones and  advances in computer vision enabled novel food recognition systems for dietary assessment. Once we recognize the category or  ingredients of the meal, we can further conduct various  health-related analysis, e.g.,  calorie intake estimation, nutrition analysis and eating habits analysis. In addition, recognizing food directly from  images is  also highly desirable for other food-related applications. Take self-service restaurants as an example, food recognition  can not only  monitor the  food consumption, but also  automatically bill the grabbed meal by the customer. Finally, for people who would like to get a better understanding of  food that they are not familiar of or they haven't even seen before, they can simply took a picture and get to know more details about it.

For these reasons, we have seen an explosion of food recognition algorithms in recent years, which are generally divided into the following four types: (1) single-label food recognition, which targets for  food images with only one food-item, (2) multi-label food recognition and detection  for  food images with multiple food-items, (3) mobile food recognition and (4) restaurant-specific food recognition.

\subsubsection{Single-label Food Recognition}
Most research works on food recognition assume that only one food item is present in one food image. Relevant works include both hand-crafted \cite{Yang-FR-CVPR2010,Hoashi2010Image,Martinel2015A} and deep \cite{Kagaya-FDR-MM2014,Wu-LMBM-MM2016,Martinel-WSR-WACV2018} representations for recognition.

There are two kinds of ways  using hand-crafted features, single type of features or the combination of different types. SIFT features~\cite{Lowe2004Distinctive}  are widely used as  visual features for food classification \cite{Wu2009Fast} \cite{Anthimopoulos2014A}. For example, \cite{Yang-FR-CVPR2010} first employed the semantic texton forest to classify all image pixels into several categories and then obtained the pairwise feature distribution as  visual features. In contrast, most methods combine different types of hand-crafted features to improve the performance of food recognition. For example, both \cite{Joutou2010A} and \cite{Hoashi2010Image} fused various types of image features including SIFT, Gabor, and color histograms via the multiple kernel learning for food recognition. For example, \cite{Nguyen2014Food}  exploited both local LBP features  and global structural information of food objects.  \cite{Martinel2015A} used  many  types of features such as Garbor, LBP and GIST,  and then exploited  a subset  to obtain the optimal ranking performance.

\begin{figure*}
\centering
\includegraphics[width=0.85\textwidth]{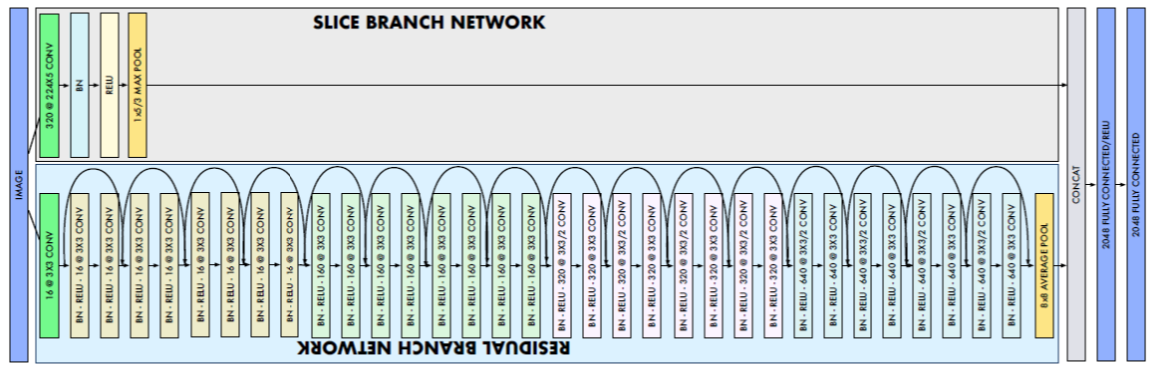}
\caption{The  WISeR architecture for food recognition~\cite{Martinel-WSR-WACV2018}}
\label{WISeR}
\end{figure*}

Recently, CNN has been  widely used for feature extraction in food recognition and achieves great performance improvement than conventional ones \cite{Krizhevsky2012ImageNet}. Different types of networks are  used in the food recognition task. For example, \cite{Kagaya-FDR-MM2014} applied the AlexNet network \cite{Krizhevsky2012ImageNet} to extract visual features for food recognition. \cite{Kawano2014Food} integrated CNN features with  conventional features as the final feature representation. \cite{Yanai-FIRDCNN-ICME2015} examined the effectiveness of the Alexnet network for food  recognition task with  pre-training and fine-tuning. \cite{Wu-LMBM-MM2016} leveraged hierarchical semantics for food recognition based on joint deep feature learning from GoogLeNet and semantic label inference. \cite{Tanno-DeepFoodCam-MM2016} used a multi-scale Network-In-Networks (NIN) \cite{Lin2013Network} to extract visual features for classification. \cite{Hassanne-FIRDCN-MM2016} fine-tuned the Inception V3\footnote{http://download.tensorflow.org/models/image/imagenet/inception-v3-2016-03-01.tar.gz.} for classifying food images. \cite{Ming2018Food} used the Resnet-50 \cite{Kaiming-Resnet-CVPR2016} to extract visual features for food recognition and nutrition analysis. \cite{Pandey2017FoodNet} proposed CNN based ensemble network architecture including the AlexNet, GoogLeNet and Resnet for food recognition.  \cite{mcallister2018combining} combined  deep residual network features with different supervised machine learning algorithms to classify food images on  diverse food image datasets.   \cite{Martinel-WSR-WACV2018} combined the  extracted visual features from the wide residual networks (WRNs) \cite{Sergey-WRN-BMVC2016} with ones  from their proposed  slice network for food recognition. Figure \ref{WISeR} shows the  WISeR architecture, which consists of two branches: a residual network branch and a slice branch network with slice convolutional layers. The residual network is to capture  general visual features while the slice branch network is to capture  specific vertical food layers. The output of two branches is fused via the concatenation and then fed to  two fully connected layers for food classification. To the best of our knowledge, the performance of this method has achieved the state-of-the-art  on  benchmark datasets. In addition, some work such as \cite{Wang-RRLMFD-ICME2015} focused on  feature fusion from different modalities including  images and associated text for food recognition.

Besides  food recognition by the  food name, food can be categorized by the cuisine. For example, \cite{Zhang-ICPF-TR2011} represented ingredients as attributes in the first layer and the cuisine as categories in the second layer, then adopted an attribute-based classifier like \cite{Lampert-DUOC-CVPR2009} for cuisine classification. \cite{Su-ARCCI-IJCPUC2014} treated the ingredients as features and constructed different classifiers to predict cuisine labels of recipes. \cite{WeiqingMin-BSC-TMM2017} utilized a multimodal deep Boltzmann machine to explore both visual and ingredient information for multi-modal recipe classification. In addition, \cite{Druck-RAP-IJCAI2013} utilized various information from the recipe, including the title, a set of ingredients and an ordered list of preparation steps for recipe attribute (such as tastes and flavors) prediction.

In addition, some work \cite{Maruyama2010Personalization, Ao2015Adapting, Horiguchi-PCFIR-TMM2018} focused on  incremental learning for food recognition.  \cite{Kaur-CWWSL-arXiv2017} augmented the deep model with noisy web food images to tackle the food recognition problem. There are also some work for food industry. For example, \cite{Russo2002A} presented a system for monitoring the work of a fastfood employee using a single static camera and  a video of an employee preparing a hamburger or sandwich is analyzed. Finally, the system produced a description of the sandwich that was prepared.  Benefiting from the large-scale food data from  social media, some works \cite{Rich-TBUASF-ICDH2016}  \cite{Barranco-PWTTW-arXiv2016} learned to recognize food image content from social media, such as Instagram and yelp.

\subsubsection{Multiple-label Food Recognition and Detection}

In real-world scenarios, there may be more than one food item in the image. The first work to  recognize multiple-food items from one food image  is proposed by \cite{Matsuda2012Recognition}. They  first detected candidate regions  and then classified them. \cite{Matsuda2012Multiple} further exploited the co-occurrence relation information between food items for recognizing multiple-food meal photos. In addition, food  detection and segmentation are widely used  for images with multiple food items.

Food detection has earlier been typically addressed as a binary classification problem, where the algorithm is simply used to distinguish whether a given image is representing food or not, namely binary food detection~\cite{Kagaya-FDR-MM2014,Ragusa2016Food}. Relevant works have considered either hand-crafted representations \cite{Kitamura2009FoodLog,miyano2012food,Farinella2015On} or deep representations \cite{Meyers-Im2Calories-ICCV2015,Kagaya2015Highly,Masashi-RMFI-ieicet19}.  Compared with hand-crafted features, an improvement is achieved via CNN based deep networks~\cite{Kagaya-FDR-MM2014}. Numerous researchers have proposed CNN based models either for feature extraction \cite{Ragusa2016Food,Aguilar2017Exploring} or for the whole recognition process \cite{Kagaya2015Highly,Singla-FnFC-MM2016}. For example, \cite{Singla-FnFC-MM2016} reported the experiments on food/non-food classification  using the GoogLeNet  network.  The best performance obtained on the public datasets with more than 15,000 images  have been reported in \cite{Aguilar2017Exploring} through the combination of the GoogLeNet for feature extraction, PCA for dimensionality reduction and SVM for classification. Different from binary food detection, \cite{Bola2017Simultaneous} recently produced a food activation map on the input image for generating proposals of the bounding boxes and then used the GoogLeNet to recognize each of  food types or food-related objects present in each bounding box. \cite{Aguilar2018Grab}  fine-tuned  the object detection algorithm  YOLOv2 \cite{Redmon2016You} for food detection with multiple-food-items. Compared with food detection,  food segmentation consists in classifying each pixel of the images representing a food. \cite{Anthimopoulos2013Segmentation} proposed a  novel method for  automatic segmentation and recognition of multi-food images, where each of the resulted segments is described by both color and texture features and classified by SVM. The latest research proposes an automatic weakly supervised method  based on CNN \cite{Shimoda2015CNN} or distinct class-specific saliency maps \cite{Shimoda2016Foodness}, respectively.  For example, \cite{Shimoda2015CNN} proposed a new region segmentation method via Region-CNN \cite{Girshick2013Rich} for food images with multiple food items.

\begin{figure*}
\centering
\includegraphics[width=0.85\textwidth]{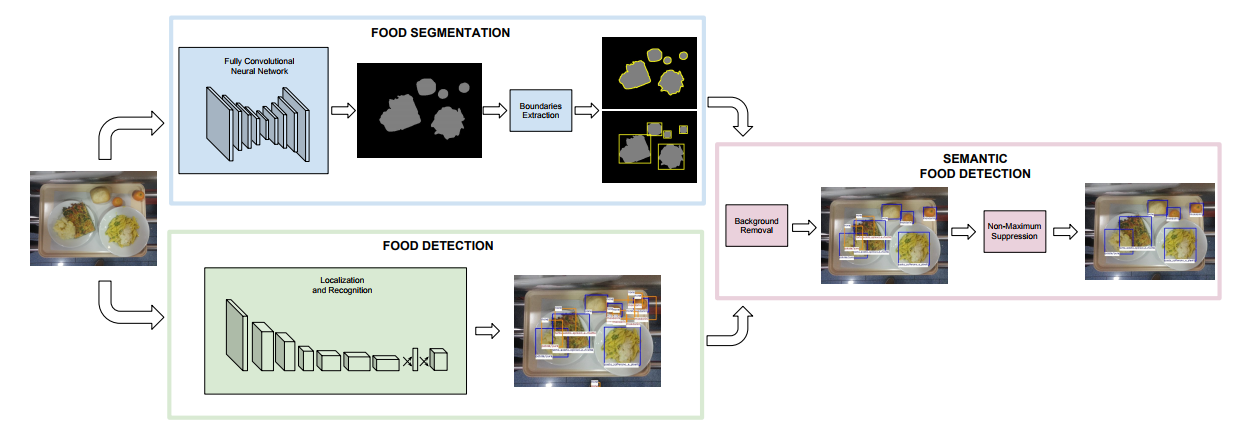}
\caption{The pipeline of  semantic food detection~\cite{Aguilar2018Grab}}
\label{SFC}
\end{figure*}

As representative work, \cite{Aguilar2018Grab} proposed a semantic food detection framework (Figure \ref{SFC}), which consists of three parts, namely food segmentation, food detection and semantic food detection. Food segmentation uses the fully CNNs \cite{Shelhamer2014Fully} to produce the binary image, and then adopts the Moore-Neighbor tracing algorithm to conduct  boundary extraction. Food detection is achieved by  retraining YOLOv2 \cite{Redmon2016You}. Semantic food detection removes  errors from object detection by combining  results of segmentation and detection to obtain final food detection results.

Besides  food recognition by food items, there are also some work on multi-label ingredient recognition. For example,  Both \cite{Chen2017Fast} and \cite{Pan2017DeepFood} used  deep learning methods  for automatic multi-class classification of food ingredients. \cite{Bola2017Food} further proposed a method for multi-label ingredient prediction via the inception v3 and Resnet-50. \cite{Zhang-ELS-CVPR2016} designed a  multi-task learning framework with the embedded ingredient structure for multi-label ingredient recognition. \cite{zhang2016multi} proposed a multi-task system that can identify dish types, food ingredients and cooking methods from food images with CNN. \cite{Zhou2016} proposed a novel approach to exploit  rich ingredient and label relationships through bipartite-graph labels, and then combined  bipartite-graph labels and CNN in a unified framework  for multi-label ingredient recognition and dish recognition.

\subsubsection{Mobile Food Recognition}
The possibility of introducing smart multimedia applications in mobile environments is gaining more and more attention, due to the rapid spreading of smart portable devices. As a consequence, there is an increasing interest on  applying food recognition to mobile environment to enable mobile food recognition. This also has other advantages of combined  built-in inertial sensors with visual food recognition to jointly monitor activities of daily living, thus providing detailed information for the dietary assessments and management \cite{Pouladzadeh2016Food}.  \cite{Zhu2011Segmentation} developed a  method to identify food items using a single image acquired from the mobile device. They firstly automatically determined  regions in an image where a particular food is located and then identified the food type based on its features including the color and texture features. \cite{Kong2011DietCam} proposed a camera phone based application DietCam, which considers food appearance from multiple perspectives for food recognition. \cite{Oliveira-MFI-PR2014} presented a semi-automatic system to recognize prepared meals which is lightweight and can be easily embedded on a camera-equipped mobile device. \cite{Rav2015Real} proposed a real-time food recognition platform combined with daily activities and energy expenditure estimation.  \cite{Kawano2015FoodCam} proposed a mobile food recognition system FoodCam to enable real-time food image recognition on a  smartphone. Deep learning offers a powerful tool to automatically produce high-level representation of complex multimedia data. Therefore, many deep learning networks, such as DenseNet \cite{Huang-Densenet-CVPR2017}, MobileNets \cite{Howard2017MobileNets} and ShuffleNets \cite{Zhang2017ShuffleNet} have been proposed to adapt to the mobile devices. Therefore, deep learning based mobile food recognition methods have been fast developed. For example, \cite{Tanno-DeepFoodCam-MM2016} extended FoodCam to DeepFoodCam by introducing the deep learning network for visual feature extraction. \cite{Pouladzadeh2017Mobile} proposed a mobile food recognition system that can recognize multiple food items in one meal, such as steak and potatoes on the same plate to further estimate the calorie and nutrition of the meal.

\subsubsection{Restaurant-specific food recognition}
\begin{figure}
\centering
 \includegraphics[width=0.70\textwidth]{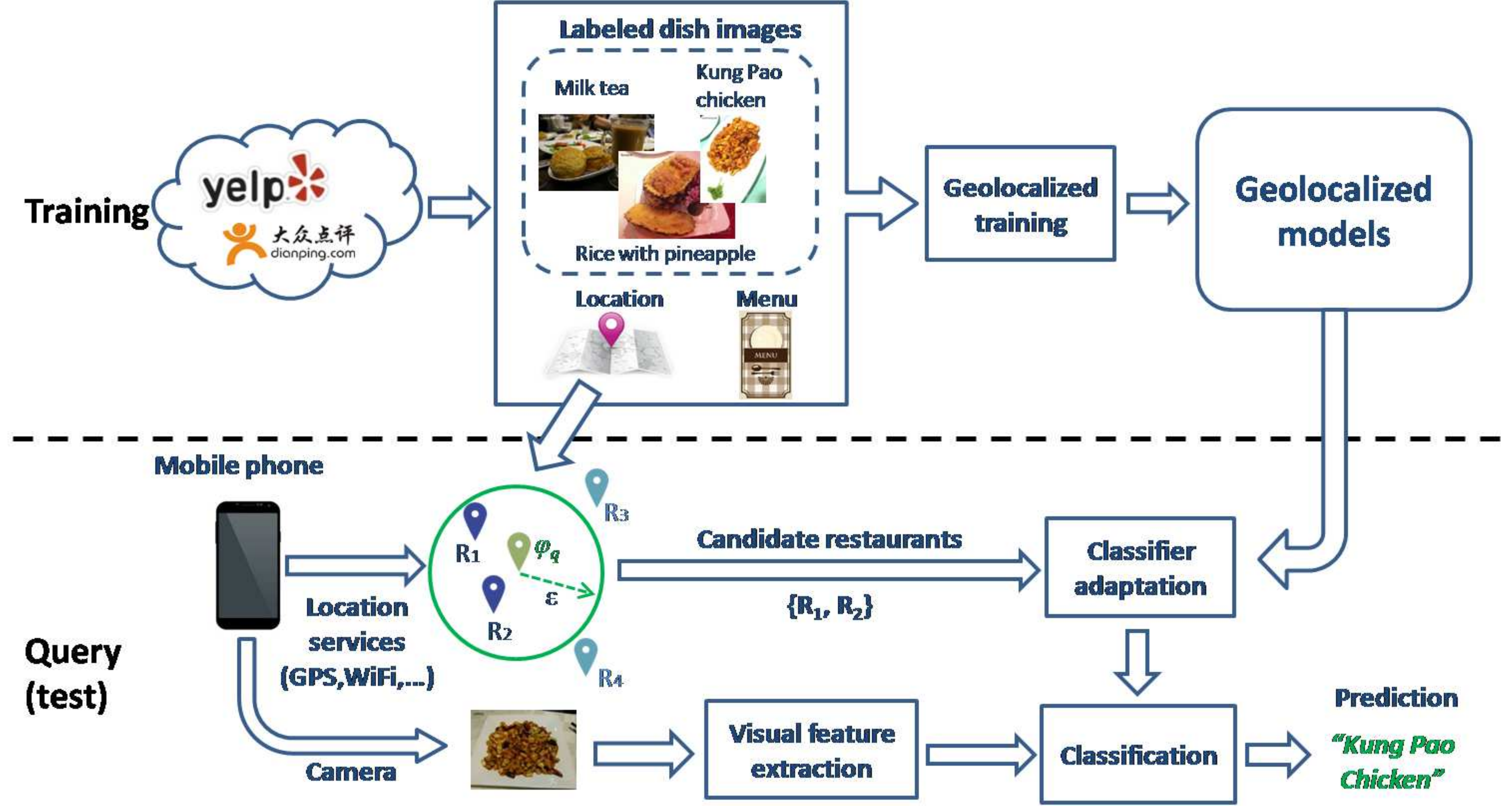}
\caption{The framework of restaurant-specific dish recognition \cite{XuRuihan-GMDR-TMM2015}}
\label{fig_GMDR}
\end{figure}
Besides  mobile food recognition, there are some recent surveys indicating that more people from many countries opt to dine out, rather than at home\footnote{http://www.whig.com/20171204/more-americans-opt-to-dine-out-rather-than-at-home\#}. Therefore, more and more works focus on the restaurant scenario for restaurant-specific food recognition. In this scenario, additional information such as the location and menu information is utilized. For example, \cite{Bettadapura-LCSAFR-WACV2015} presented a method for automatically recognizing food  in restaurants leveraging location sensor data and various hand-crafted visual features. \cite{XuRuihan-GMDR-TMM2015} proposed a framework incorporating discriminative classification in geolocalized settings and introduced the concept of geolocalized models, where  DeCAF deep features and  restaurant location information are utilized. As shown in Figure \ref{fig_GMDR},  they first constructed a database of restaurants including geographical locations and menus and food images of  corresponding dishes. Then, the geolocalized models are trained with these dish images, where each model is related to a particular geolocation. During the test time, the particular geolocation of the query defines a neighborhood with some candidate restaurants. For each query, corresponding geolocalized models are selected and combined into a new classifier adapted to the query. In addition, \cite{Herranz-PMFIR-ICME2015,Herranz-MRCFR-TMM2017} proposed a probabilistic graph model to  connect dishes, restaurants and locations for food recognition. \cite{Wang2016Where} proposed a multi-task convolutional neural network  for simultaneous dish and restaurant recognition from food images. \cite{Aguilar2018Grab} conducted automatic food tray analysis in canteens and restaurants by food detection and segmentation  based on CNNs for smart restaurants. \cite{Marc-WWE-ECCVW18} combined CNNs and Recurrent Neural Networks (RNNs)  to determine its correct menu item corresponding to the restaurant given an image as the input.

Table \ref{recognition_1} and Table \ref{recognition_2} provides an overview of these approaches with respect to visual features, additional information and recognition type. The classifiers  most methods adopt are SVM or Softmax. Table \ref{recognition_performance} shows an overview of current performance comparison on  benchmark datasets.

\begin{table*}[htbp]
\caption{Summary of Food Recognition Using Conventional Visual Features}
\begin{center}
    \begin{tabular}{|c|c|c|c|}
        \hline
        Reference & Visual Features & Additional Information & Recognition Type \\
        \hline
        \cite{Bolle1996VeggieVision}&Texture, Color&-&Food  recognition\\
        \cite{Puri2009Recognition}&Color, Textures &-& Mobile food  recognition \\
        \cite{Wu2009Fast}&SIFT&-& Food  recognition\\
        \cite{Hoashi2010Image}&\tabincell{c}{SIFT,Color\\ Texture, HoG} &-& Food  recognition \\
        \cite{Joutou2010A}&SIFT,Color, Texture &-& Food  recognition \\
        \cite{Yang-FR-CVPR2010} &\tabincell{c}{Pairwise Local Features\\ Joint Pairwise Local Features} &-& Food  recognition \\
        \cite{Zong2010On} & SIFT, Texture & - & Food  recognition\\
         \cite{Bosch2011COMBINING} & SIFT, Color, Texture &-& Food recognition \\
         \cite{Zhang-ICPF-TR2011} &Color, Texture  &-& Cuisine classification\\
         \cite{Matsuda2012Multiple} &SIFT, Color, HoG, Texture &-& Food recognition \\
         \cite{Matsuda2012Recognition} &\tabincell{c}{SIFT, Color\\ HoG, Texture} &-& Food  recognition \\
        \cite{Farinella-CFI-ICIP2014} &Texture &-& Food  recognition \\
        \cite{Nguyen2014Food} &SIFT, Texture, Shape &-& Food  recognition\\
        \cite{Anthimopoulos2014A} & SIFT, Color & - & Food  recognition \\
        \cite{Oliveira-MFI-PR2014} &Color, Texture &-& Mobile food  recognition\\
        \cite{Kawano-FoodCam-MM2014} &HoG, Color &-& Mobile food  recognition\\
        \cite{Farinella2015On} & SIFT, Texture, Color & - & Food  recognition \\
        \cite{Martinel2015A} & Color, Shape, Texture &-&  Food  recognition \\
        \cite{Bettadapura-LCSAFR-WACV2015} &SIFT, Color &Location \& Menu& \tabincell{c}{Restaurant-specific \\food  recognition} \\
        \cite{Farinella2015Food} &SIFT, SPIN &-& Food  recognition \\
        \cite{Kawano2015FoodCam} &SIFT, Color, HoG &-& Mobile food  recognition\\
        \cite{Rav2015Real} &HoG, Texture, Color&-& Mobile food  recognition \\
       \cite{Martinel2016A} &SIFT, Color, Shape, Texture &-& Food  recognition\\
        \cite{He2017DietCam} & Texture &-& Food  recognition\\
        \cite{zheng2017food} &SIFT, Color &-& Food  recognition \\
        \hline
    \end{tabular}
    \end{center}
    \label{recognition_1}
\end{table*}

\begin{table*}[htbp]
\caption{Summary of Food Recognition Using Deep Visual Features}
\begin{center}
    \begin{tabular}{|c|c|c|c|}
        \hline
        Reference & Visual Features & Additional Information & Recognition Type \\
        \hline
        \cite{Kawano2014Food} &HoG, Color, CNN&-& Food recognition\\
        \cite{Simonyan-VDCN-arXiv2014}& VGG &-& Food  recognition \\
        \cite{Kagaya-FDR-MM2014}& AlexNet &-& Food  recognition \\
        \cite{Ao2015Adapting}&GoogleNet&-&Food  recognition  \\
        \cite{Yanai-FIRDCNN-ICME2015}& AlexNet &-& Food  recognition \\
        \cite{Christodoulidis-FRDA-ICIAP2015}& CNN &-& Food  recognition\\
        \cite{Wang-RRLMFD-ICME2015}&VGG&Text&Recipe recognition \\
        \cite{XuRuihan-GMDR-TMM2015}&DeCAF &Location& \tabincell{c}{Restaurant-specific \\food  recognition} \\
        \cite{Herranz-PMFIR-ICME2015}&DeCAF &Location& \tabincell{c}{Restaurant-specific \\food  recognition}\\
        \cite{Herruzo2016Can} &GoogleNet&-& Food  recognition \\
         \cite{Wang2016Where}&CNN&Location& \tabincell{c}{Restaurant-specific \\food  recognition}\\
        \cite{Singla-FnFC-MM2016}&GoogleNet&-&Food  recognition \\
         \cite{Ragusa2016Food}& AlexNet, VGG, NIN&-&Food  recognition \\
        \cite{Wu-LMBM-MM2016}&GoogleNet&-&Food  recognition \\
        \cite{Ciocca2016Food}& AlexNet &-& Food  recognition \\
        \cite{Liu2016DeepFood}&Inception  &-& Food  recognition\\
        \cite{Hassanne-FIRDCN-MM2016}&Inception  &-& Food  recognition\\
        \cite{Tanno-DeepFoodCam-MM2016}&Network In Network&-&  Mobile food  recognition\\
        \cite{Herranz-MRCFR-TMM2017}& AlexNet &Location \& Menu&\tabincell{c}{Restaurant-specific \\food  recognition}\\
        \cite{Bola2017Simultaneous}&GoogleNet&-&Food  recognition \\
        \cite{Pandey2017FoodNet}& \tabincell{c}{ AlexNet, GoogLeNet \\ ResNet}& -& Food  recognition\\
        \cite{Chen2017ChineseFoodNet}&\tabincell{c}{ ResNet-152, DenseNet\\VGG-19}&-&Food  recognition\\
        \cite{termritthikun2017nu}&NUInNet &-& Food  recognition\\
        \cite{Kaur-CWWSL-arXiv2017}&Inception-ResNet&-&Food  recognition\\
        \cite{Pan2017DeepFood}&\tabincell{c}{ AlexNet, CafffeNet\\RestNet-50}&-&Ingredient classification \\
        \cite{Aguilar2017Food}&\tabincell{c}{InceptionV3, GoogLeNet\\ ResNet-50}&-& Food  recognition \\
        \cite{mcallister2018combining}&ResNet-152, GoogleNet&-& Food  recognition \\
        \cite{Ming2018Food}&ResNet-50&-&  Mobile food  recognition\\
        \cite{Martinel-WSR-WACV2018}& WISeR& -& Food  recognition\\
        \hline
     \end{tabular}
    \end{center}
     \label{recognition_2}
\end{table*}

\begin{table*}[htbp]
\caption{Performance Comparison on the Accuracy in Three Benchmark Datasets (\%).}
\begin{center}
    \begin{tabular}{|c|c|c|c|}
        \hline
        Reference & UECFood100 &  UECFood256 & ETHZ Food-101 \\
        \hline
        \cite{Kawano2014Food}&72.26&-&-\\
        \cite{Kawano-FoodCam-MM2014}&-&50.10&-\\
        \cite{Rav2015Real}&53.35&-&- \\
         \cite{Martinel2015A} &80.33&-& -\\
         \cite{Yanai-FIRDCNN-ICME2015}&78.77&67.57&70.41\\
         \cite{Ao2015Adapting}&-&-& 78.11 \\
         \cite{Wu-LMBM-MM2016}&-&-& 72.11 \\
        \cite{Liu2016DeepFood}&76.30&54.70&77.40\\
        \cite{Martinel2016A} &84.31&-&55.89\\
        \cite{Hassanne-FIRDCN-MM2016}&81.45&76.17&88.28\\
        \cite{zheng2017food} &70.84&-&- \\
        \cite{Bola2017Simultaneous}&-&63.16&79.20\\
        \cite{Aguilar2017Food}&-&-&86.71\\
        \cite{Pandey2017FoodNet}&-&-&72.12\\
        \cite{mcallister2018combining}&-&-&64.98\\
        \cite{Martinel-WSR-WACV2018} &\textbf{89.58}&\textbf{83.15}&\textbf{90.27} \\
        \hline
     \end{tabular}
    \end{center}
     \label{recognition_performance}
\end{table*}

\subsection{Retrieval}
These massive amounts of data shared on various sites allow gathering food-related data such as text recipes, images, videos and user preference. A food-relevant retrieval engine is necessary to obtain what we need. In health-oriented applications, predicting nutrition content and calorie information from food images requires fine-grained ingredient recognition. However, directly recognizing ingredients is  sometimes challenging, since ingredients from prepared foods  are mixed and stirred. In this case, we can retrieve recipes based on  the image query, namely cross-modal retrieval. There are  other  advantages of having recipes. For example, recipe recommendation will help users  cook a particular dish what he wants. In addition, these recipes provide other rich information, e.g., ingredients, macronutrient composition and cooking methods, which can enable the  estimation of food calorie and nutrition facts.

According to retrieval types, food-relevant retrieval consists of three types: visual food retrieval, recipe retrieval and cross-modal recipe-image retrieval. For food image retrieval, \cite{Kitamura2009FoodLog} proposed a  FoodLog system, which can distinguish food images from other images to retrieve personal food images for food balance estimation via the combination of BoF visual features and SVM.  Compared with \cite{Kitamura2009FoodLog}, \cite{Aizawa2014Comparative} improved the  food image retrieval system by supporting both image-based and text-based query. \cite{barlacchi2016appetitoso} introduced a  search engine for restaurant retrieval based on dishes a user would like to taste rather than using the name of food facilities or their general categories. \cite{Farinella2016Retrieval} conducted the food image retrieval by comparing the images through similarity measures, where the food images are represented as vectors through the combination of different types of features, such as  SIFT and Bag of Textons. \cite{ciocca2017learning} adopted CNN-based features for food image retrieval. For the recipe retrieval, \cite{Wang-SSMCR-WWW2008} investigated the underlying features of Chinese recipes. Based on workflow-like cooking procedures, they model recipes as graphs and  further propose a novel similarity measurement based on the frequent patterns, and devise an effective filtering algorithm  to support efficient on-line searching. Recently, \cite{Chang-RecipeScape-CHI2018} proposed an interactive system RecipeScape to  analyze multiple recipes for one dish. They changed the recipe instruction into a tree-structure representation for recipe similarity calculation. \cite{Xie2011A} further jointly utilized various features such as cooking flow features, eating features and nutrition features to create a hybrid semantic item model  for recipe search.

Besides food/recipe retrieval, there are some researches on cross-modal recipe-image retrieval. For example, \cite{Chen-DIRCRR-MM2016} proposed a multi-task deep learning  architecture for simultaneous  ingredient and food recognition. The learnt visual features and semantic attributes of ingredients  are then used for recipe retrieval given dish pictures.  \cite{Chen-CMRR-MMM2017} introduced a stacked attention network to learn  joint space from images and recipes for cross-modal retrieval. \cite{Jing-CMR-MM2017} exploited rich food attributes for cross-modal recipe retrieval. \cite{WeiqingMin-BSC-TMM2017} utilized a  multi-modal Deep Boltzmann Machine  for  recipe-image retrieval. \cite{Salvador-LCME-CVPR2017} developed a hybrid neural network architecture, which jointly learned shared space via  image and recipe embedding  for cross-modal  image-recipe retrieval, where visual features are learned by  CNN  while  recipe features are learned by LSTM. \cite{Carvalho2018SIGIR} extended \cite{Salvador-LCME-CVPR2017} by providing a double-triplet strategy to jointly express  both the retrieval loss  and the classification one for cross-modal retrieval.\cite{Wang-LCME-CVPR2019,Zhu-R2GAN-CVPR2019} further introduced adversarial networks to  impose the modality alignment for cross-modal retrieval. \cite{Salvador-IC-CVPR2019} proposed a new architecture for ingredient prediction that exploits co-dependencies among ingredients without imposing order for generating cooking instructions from an image and its ingredients.

\begin{figure*}
\centering
\includegraphics[width=0.70\textwidth]{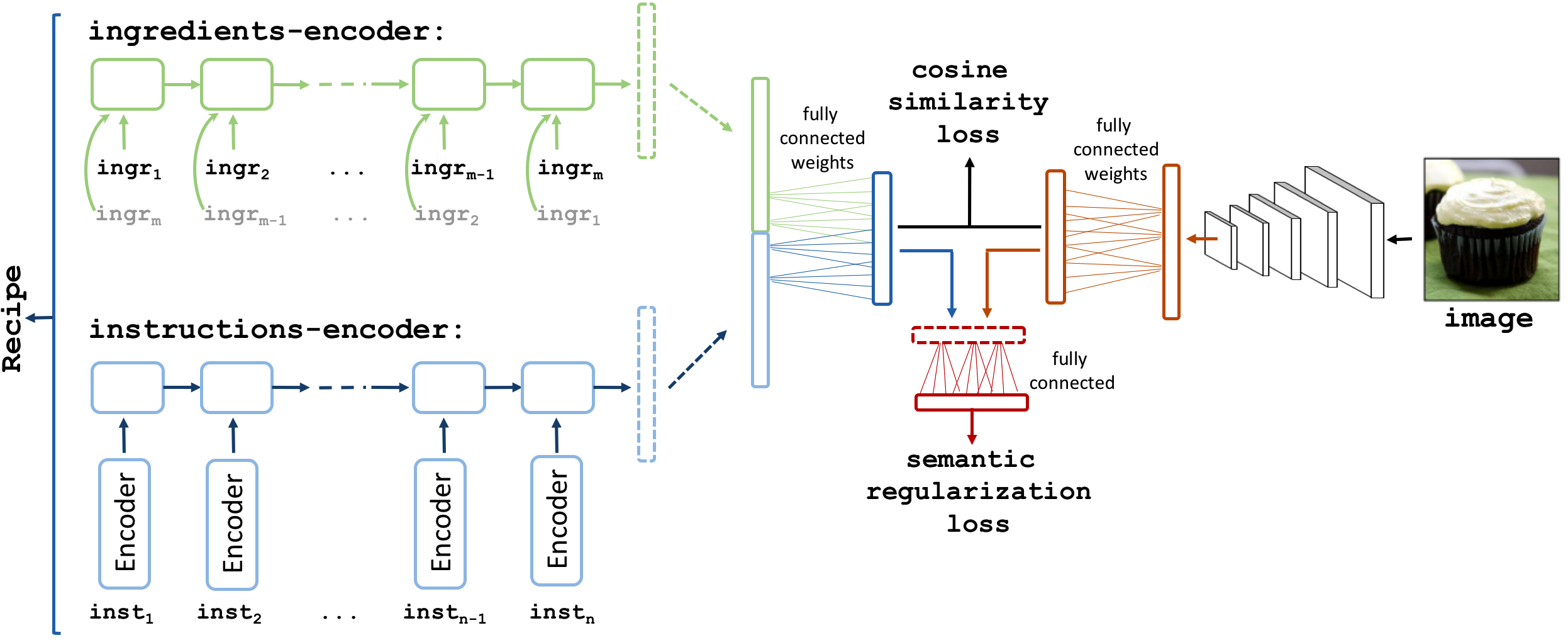}
\caption{The proposed im2recipe architecture for cross-modal recipe-image retrieval~\cite{Salvador-LCME-CVPR2017}}
\label{im2recipe}
\end{figure*}

As  representative work,  Figure \ref{im2recipe} shows the proposed   joint embedding model. There are mainly two components for a recipe:  ingredients and cooking instructions. For  ingredients, they first extracted the ingredient name using bi-directional LSTM \cite{Schuster1997Bidirectional}. Then each ingredient name is represented via the  word2vec model \cite{Mikolov2013Efficient}. Finally, a bidirectional LSTM model is again used to encode these ingredients to the feature representation.  For the cooking instruction,  they utilized the  LSTM model to encode it to a  fixed-length feature representation.  These two kinds of representations are concatenated  to  the final recipe representation. For the image representation, two  deep convolutional networks, namely VGG-16 and Resnet-50 models are adopted to extract  visual features. Additional semantic regularization on the embedding is further introduced to improve joint embedding.

Table \ref{retrieval_table}  provides a summary of  retrieval approaches with respect to  features, dataset and tasks.

\begin{table*}[htbp]
\caption{Summary of Main Retrieval Methods}
\begin{center}
    \begin{tabular}{|c|c|c|c|c|}
        \hline
        \multirow{2}*{Reference}&
        \multicolumn{2}{c|}{Data type}&
        \multirow{2}*{ Dataset Name}&
        \multirow{2}*{ Task }\\
         \cline{2-3}&Image&Text & \multicolumn{1}{c|}{}&\multicolumn{1}{c|}{}\\
        \hline
        \cite{Wang-SSMCR-WWW2008}&-&Cooking graph&Cooking graph database&Recipe retrieval \\
        \cite{Kitamura2009FoodLog} &Food images&-&FoodLog&Food retrieval\\
        \cite{Xie2011A}&-&Cooking graph& - &Recipe retrieval\\
        \cite{barlacchi2016appetitoso}&-&Dish name \& Ingredients&\tabincell{c}{Food Taste Knowledge\\ Base (FKB)}& Recipe retrieval\\
        \cite{Farinella2016Retrieval}&Food images&-&UNICT-FD1200&Food retrieval\\
        \cite{Chen-DIRCRR-MM2016}&Food images &Ingredients&VIREO Food-172&\tabincell{c}{Cross-modal\\ retrieval}\\
        \cite{Chen-CMRR-MMM2017}& Food images& Ingredients&-&\tabincell{c}{Cross-modal\\ retrieval}\\
        \cite{Jing-CMR-MM2017}&Food images&Ingredients&-&\tabincell{c}{Cross-modal\\ retrieval}\\
        \cite{Salvador-LCME-CVPR2017}&Food images& Ingredients \& Instructions &Recipe 1M &\tabincell{c}{Cross-modal\\ retrieval}\\
        \cite{WeiqingMin-BSC-TMM2017} &Food images &Ingredients \& Attributes&Yummly-28K& \tabincell{c}{Cross-modal\\ retrieval}\\
        \cite{ciocca2017learning}&Food images&-&Food524DB&Food retrieval \\
        \cite{Carvalho2018SIGIR}&Food images& Ingredients \& Instructions&Recipe 1M &\tabincell{c}{Cross-modal\\ retrieval}\\
    \cite{Wang-LCME-CVPR2019}&Food images& Ingredients \& Instructions&Recipe 1M &\tabincell{c}{Cross-modal\\ retrieval}\\
    \cite{Zhu-R2GAN-CVPR2019}&Food images& Ingredients \& Instructions&Recipe 1M &\tabincell{c}{Cross-modal\\ retrieval}\\
    \cite{Salvador-IC-CVPR2019}&Food images& Ingredients \& Instructions&Recipe 1M &\tabincell{c}{Cross-modal\\ retrieval}\\
        \hline
     \end{tabular}
    \end{center}
\label{retrieval_table}
\end{table*}

\subsection{Recommendation}
Food recommendation is an important domain  for both individuals and society.  Different from other types of recommendation system, food recommendation involves more complex, multi-faceted and other context-dependent information (e.g. life-style preferences and culture) in predicting what people would like to eat \cite{Trattner2017Food}.  Taking all these factors into consideration, various recommendation methods are proposed. According to the recent survey \cite{Trattner2017Food}, food recommendation  consists of four types, namely content-based food recommendation \cite{Freyne-IFP-ICIUI2010}, collaborative filtering-based food recommendation, hybrid food recommendation \cite{Ge-UTLF-ICDH2015}, context-aware food recommendation \cite{Cheng-FR-UMAP2017,Zhang-EDPRR-WWW2016} and health-aware food recommendation \cite{NagPSLWJ17}\cite{Yang2017Yum,Schafer-Hanna-DH2017,Achananuparp2016Extracting,Trattner-IHISR-WWW2017}.

For content-based food recommendation, recipe oriented recommendation has been extensively studied. For example, \cite{Freyne-IFP-ICIUI2010} made recommendations by breaking recipes down into individual ingredients and scoring based on the ingredients contained within recipes, which users had rated positively. In such methods, different methods for recipe based content representation are adopted, including topic model based  representation \cite{Nedovic-LRIS-CWC2013,Kusmierczyk-FRTS-CIKM2016}, structure based representation \cite{Jermsurawong-PSCR-EMNLP2015,Kiddon-LIGIR-Thesis2016} and multi-modal representation with various attributes \cite{Min-DRA-MM2017,Min-YAWYE-TMM2018}.

For collaborative filtering-based methods, classic Singular Value Decomposition (SVD) \cite{Harvey2013You} and Matrix Factorization (MF) \cite{Ge-UTLF-ICDH2015}  have been used widely for recommendation. For example, \cite{Ge-UTLF-ICDH2015} utilized a MF approach for food recommender systems
that fuses ratings information and user supplied tags to achieve significantly better prediction accuracy than content-based and standard matrix factorization baselines. In addition, other methods such as Latent Dirichlet Allocation and Weighted matrix factorization are also used for food recommendation \cite{Trattner-IHISR-WWW2017}.

For context-aware approaches, numerous exploratory data analysis have demonstrated that rich context such as gender, time, hobbies, location and cultural aspects is important in food recommendation. For example,  \cite{Cheng-FR-UMAP2017} conducted context-aware food recommenders, which are created by filtering users and items according to  relevant context factors.  In addition, exploring  other factors such as the culinary cultures can also help for context-aware food recommendation. For example, \cite{Ahn-FNFP-SciRe2011} constructed a data-driven flavor network relating ingredients to discover the patterns of  ingredient combinations. \cite{Golder2011Diurnal} has discovered  clear, universal rhythmic patterns regarding work, sleep and eating from millions of public Twitter messages. Similar temporal and spatial patterns can be found by analyzing online recipe websites \cite{Wagner2014The} \cite{Kusmierczyk2015Temporal}. \cite{Silva-YAWYE-Arxiv2014} proposed a new method to identify cultural boundaries and similarities across populations at different scales based on the analysis of Foursquare check-ins. Such cultural analysis and understanding from recipes and social media can help us  develop recommendation mechanisms considering the cultural characterization of specific urban areas. In addition, we can also discover user's food preference \cite{Kusmierczyk2015Temporality} for context-aware food recommendation.

Health-aware food recommendation is unique for food recommendation. Incorporating health into the recommendation has largely been a recent focus \cite{Ge-HFRS-Rec2015,Nag2017Live,Markus-Recipe-ICWSM2018}. For example, \cite{Ge-HFRS-Rec2015} incorporates nutritional aspects into the recommendation approach  based on a so-called ''calorie balance function''. \cite{Nag2017Live} proposed a live personalized nutrition recommendation system, which can  efficiently calculate which items are healthiest and re-rank
and filter results to users based on their personalized health data streams and environmental context. Recently, \cite{Markus-Recipe-ICWSM2018} built a model using different kinds of features  from a recipe's title, ingredient list and cooking directions, popularity indicators such as the number of ratings and the user comments and visual  features to  estimate the healthiness of  recipes for  health-aware recipe recommendation. \cite{thanh2017healthy} proposed a method for analyzing  food and drink consumption patterns on Instagram.

In addition, \cite{Maruyama-RMRR-MM2012} proposed a mobile recommendation system, which  recognized food ingredients and recommended recipes related to these recognized ingredients to users in a real-time way. \cite{Fadhil2017} addressed some challenges of chatbot application for meal recommendation. \cite{Phanich2010Food} proposed a food recommendation system for diabetic patients  using food clustering analysis to recommend  proper substituted foods in the context of nutrition and food characteristic. \cite{Rehman2017Diet} used  the ant colony algorithm to generate optimal food list and recommended suitable foods for patients based on users' pathological reports. Several studies in the field of nutrition science have shown that proper nutrition and health labels help people to make better food choices \cite{Sonnenberg2013A,Elbel2011Child}. There are also some works for restaurant recommendation \cite{Zhang-EDPRR-WWW2016} or food truck recommendation \cite{Rivolli2017Food}. Besides the above-mentioned references, we refer readers to one recent survey of food recommendation in \cite{Trattner2017Food,Theodoridis-AINRS-PETRA2019}.

\subsection{Prediction and Monitoring}
Online social networks (e.g., Twitter, Instagram and Facebook) with  billions of users and shared large-scale food data, have become  rich sources  to conduct food-related  prediction and monitoring.  Prediction and monitoring from the social media can provide various health-relevant information to enable further decision, such as predicting recipe popularity \cite{Sanjo2017Recipe}, the national obesity, diabetes statistics \cite{Abbar2015You}  and monitoring the public health \cite{Capurro2014The}.

Food-related prediction in social media has gained more attention \cite{Abbar2015You,Wagner2014The,West2013From,Kusmierczyk2015Temporality}.  For instance, \cite{Ma2015Will} predicted the income from the preference for spicy foods. \cite{Mejova-FoodPorn-ICDH2015} analyzed large scale images on Instagram  to study food consumption patterns from the Unite States. \cite{Abbar2015You} used  daily tweets of users about food from Twitter to predict the national obesity and diabetes statistics.   \cite{Fried2014Analyzing}  collected a large corpus of food-related tweets from Twitter and used them to predict latent population characteristics such as  geographic location of authors, overweight and diabetes rates.  \cite{De-CDCN-CSCW2016} proposed a simple ingredient matching method to estimate nutritional properties of food posts on Instagram, making use of the USDA National Nutrient Database for the matching process. \cite{Sanjo2017Recipe} predicted the recipe popularity by fusing their multimodal features including visual and semantic features extracted from the deep network. In addition, some work such as \cite{Kusmierczyk2016Understanding} predicted online food production patterns from online food community.

Recently, using the social media for monitoring public health has received more attention \cite{Capurro2014The}. In the early years, in order to conduct large-scale dietary studies, we should use questionnaires and food diaries to keep track of participants' daily activities, which can be inaccurate and expensive.  Alternatively, social media such as Twitter and Instagram  provides its users with a way of recording their daily lives, such as  dietary choices.  Some studies have adopted data-driven  approaches to analyze the food consumption on massive scale from these records in the social media \cite{Silva-YAWYE-Arxiv2014,Culotta2014Estimating,Ofli-Saki-WWW2017}. For example, \cite{Mejova2016Fetishizing} exploited 10 million  posts from 1.7 million users on Instagram to capture  global use of the popular \#foodporn hashtag.  \cite{Sadilek2017Deploying} prevented the Foodborne illness by  mining the data in the social media. They applied the machine learning method to Twitter data and developed a system that automatically detected venues likely to pose a public health hazard. \cite{PayamKarisani2018} presented a new method  to detect personal health mentions in  Twitter.

\section{CHALLENGES}\label{section_CHA}
Food computing has received more and more attention in the last few years for its wide applications. Thus, it is extremely important to discuss existing challenges that form the major obstacles to current progress. This section presents key unresolved issues.
\subsection{Food Image Recognition}
Robust and accurate food image recognition is very essential for  various health-oriented applications, such as food calorie estimation, food journaling and  automatic dietary management.  However, it is very challenging for the following three reasons: (1) Food images have their own distinctive properties. They don't have any distinctive spatial layout. Although some food categories such as fruits, hamburgers and pizzas have regular shapes, many food dishes have deformable food appearance and are thus  lack of rigid structures. Ingredients can be the constituent part of food. However, ingredients from many types of food images are distributed randomly in a plate. Other factors, such as cooking methods also affect the appearance of food ingredients. This makes the task different from  other ones like scene recognition, where we can always find some distinctive features such as buildings and trees. Therefore, simply borrowing the methods from object or scene recognition is hard to achieve satisfactory recognition results, especially for real-world applications, not mention to images with multiple-item meals. (2) Food image recognition belongs to fine-grained classification. Similarly, food image recognition encounters the same problem as the fine-grained classification, such as  subtle differences among different food categories. However, we can not simply directly use existing fine-grained classification methods, such as \cite{Fu-LC2SB-CVPR2017} for food image recognition. The reason is that existing fine-grained categorization methods aim to distinguish between different breeds or species.  They generally first discover the fixed semantic parts, and then concatenate the features from both  global object and semantic parts as the final representation. Such representation includes not only  global features but also more discriminative local features. For example, in the bird classification, some semantic parts, such as  head and breast  should be first localized. However, the concepts of common semantic parts do not exist in food images.  Therefore, we should design a new fine-grained categorization paradigm, which is suitable for food recognition. (3) There is lack of large-scale benchmark food images  with many categories. In the computer vision, the release of  large-scale ImageNet dataset with the Wordnet ontology  has greatly further the development of object recognition \cite{Krizhevsky2012ImageNet}. Similarly, the large-scale food dataset is required. There are indeed some benchmark food datasets, such as Food101 \cite{Bossard-Food101-ECCV2014} and UEC Food256 \cite{Kawano-FoodCam-MM2014}. However, the categories and number  of these datasets are not big enough  compared with the ImageNet. In addition, food-oriented dataset construction has its particular challenges.  For example, because of the region difference, there are probably several different names for the same dish. Similarly,  some dishes are labeled with the same dish name, but actually belong to different dishes with different ingredients. This means that it is harder to build a standard  ontology according to the dish name like the Wordnet.

\subsection{Vision based Dietary  Management System}
With the fast development of computer vision and machine learning,  more and more dietary  management systems resort to vision-based methods. For example, \cite{Meyers-Im2Calories-ICCV2015} from Google proposed a system Im2Calories, which can recognize  ingredients of the meal from one food image and then predict its  calorie account. \cite{Beijbom-MeMa-WACV2015} from Microsoft and University of California  presented a computer vision system  for automatically logging the food and calorie intake from food images in the restaurant scenario. However, existing dietary management systems are far from  perfect and practical. The reasons derive from two-fold: (1) existing food recognition methods are robust to only few and standard dishes. In  real-world scenarios, there are thousands of food categories to recognize. Many types of food images do not in the training set. As a result, the system  fails to recognize the food, and then the estimated amount of calories is incorrect. In addition, most existing food recognition methods are not specifically for food images and thus have  unsatisfactory recognition performance.  (2) Even we recognize the food and localize the food region, we next should estimate the food volume. It is still hard to accurately estimate the volume from one image. Probably we can add the interaction to alleviate these problems, which conversely affect the user experience. Therefore, we should simultaneously solve the above-mentioned problems to enable a robust vision based dietary management system, which is harder to achieve.

\subsection{Multiple-Network oriented food data fusion and mining}
During the past decade, the influence of social network services on people's daily life has sharply increased. Many users  participate in different social networks. For example, one user may share food photos in Instagram, upload the recipe  to the twitter and perform check-ins in Foursquare. In order to completely predict the health and wellness to deliver better healthcare, the first step is to effectively combine and integrate these food-related multi-modal signals from different social networks. However, the unbalanced data distributions in different networks and different accounts from different networks for each user make the effective fusion more challenging. Most of  food-relevant works mentioned previously use only one data source. They may not be enough to gain deeper insights and more complete knowledge from multi-source social media data. Furthermore, besides the social network, there are other types of networks, such as  mobile networks and IoT.  Therefore, we can obtain  diverse  signals from these different networks. For example, Fitocracy and MyFitnessPal provide the exercise semantics (i.e., sports activity type). Endomondo  can be considered as a rich source of sequential data from wearable sensors and wellness-related ground truth. These mobile devices usually include rich multidimensional context information, such  as altitude, longitude, latitude and time. Computing the user's lifestyles needs further integrate these heterogeneous signals  in a unified way. To the best of our knowledge, there are few publicly works towards it. Multimodal fusion still faces other challenges. For example, it is difficult to build one model that can exploit both shared and complementary information. In addition, not all the data sources will be helpful for certain food-related  tasks in some cases. Among all these fused data sources, picking the useful ones is not an easy task.


\subsection{Health-aware Personalized Food Recommendation}

Existing methods \cite{Elahi2017User,Harvey2017Exploiting} mainly refer to the trade-off for most users between recommending the user what he/she wants and what is nutritionally appropriate, where the healthness of the recipe can be predicted based on multiple cues, such as ingredients and images. However, there are many other factors to  make health-aware personalized food recommendation challenging, such as complex, multi-faceted, information (e.g., the temporal and spatial context, culture, gender and user preference). Each person  is unique and the physical state of each person is different at different moments. To enable more accurate food recommendation, we should monitor their  wellness constantly. Although  some works \cite{Farseev-TFI-TIST2017} integrated the data from wearable devices and several social networks to learn the wellness profile, the heterogeneous modality fusion is still difficult. Therefore, when developing health-aware personalized food recommendation systems, there are additional issues  to consider, which do not arise in other recommendation domains. These include that users may have many constrained needs, such as allergies or life-style preferences, the desire to eat only fruit or vegetarian food. In such cases, existing methods work not well.

\subsection{Food Computing for Food Science}
Food computing is an inherently multidisciplinary field and its progress is predominantly dependent on support, knowledge and advances in closely related fields,
such as food science, biology, gastronomy, neuroscience and computer science. As the performance of contemporary vision systems such as food image recognition is still far from perfect. Further investigations into the mechanisms of human perception  on the visual food may be a crucially important step in gaining invaluable insights and relevant knowledge that can potentially inspire the better design of the dietary  management. For example, most existing food computing methods mainly focus on the conventional multimodal data analysis and mining. However, food science involves multiple subdisciplines, such as food chemistry and food microbiology. We should cope with new data types (e.g., the chemical forms and the molecules structure in food) and new tasks (such as immunogenic epitopes detection from the wheat). Therefore, current food computing methods must be adapted or even re-designed to  handle these new data and new tasks. For example, how to design a multimedia feature learning method to represent new data type, such as special chemical forms or the molecules structure in food? How to design novel food computing methods, which target for new tasks, such as ingredient recognition in the food  engineering environment? How to use the food computing method to  detect various food-borne illnesses in the food quantity control?


\section{FUTURE DIRECTIONS}\label{section_FD}
As mentioned earlier, considerable effort will be required in the future to tackle the challenges and open issues with  food computing. Several future directions and solutions are listed as follows.

\subsection{Large-scale Standard Food Dataset Construction}
Like ImageNet  for general objects in the computer vision, a large-scale ontology of ImageNet-level  food images is also a critical resource for developing advanced, large-scale content-based food image search, classification and  understanding algorithms, as well as for providing critical training and benchmark data for such algorithms. To construct the large-scale food dataset,  a feasible method is to combine  food image crawling from the social media and  manual annotation from the crowd-sourcing platform AMT. In addition, we should consider the geographical distribution of food images, such as different cuisines, to cover the whole world. Each region has their own special cuisines and dishes, there is no food experts to master all the dishes. Therefore, the construction of the large-scale food dataset also should need joint efforts of scientists all over the world.
\subsection{Large-scale Robust Food Recognition System}
Vision-based food system is very fundamental to various real-world applications, such as the  dietary  assessment and management system. The first priority is to develop a large-scale robust food recognition system. In recent years, deep learning approaches such as CNNs \cite{Krizhevsky2012ImageNet} and their variants (e.g., the VGG network \cite{Szegedy2015Going}, ResNet \cite{Kaiming-Resnet-CVPR2016} and DenseNet \cite{Huang-Densenet-CVPR2017}), have provided us with great opportunities to achieve this goal. Deep learning has the advantage of learning more abstract patterns progressively and automatically from raw image pixels in a multiplelayer architecture than using hand-engineered features. There are indeed some efforts for this direction. For example, \cite{Martinel-WSR-WACV2018} proposed  a slice convolution network to capture  vertical food structure, and combined   visual features from the general deep network to achieve the state-of-the art performance. We believe there are other special food structures and properties to explore. If we design the deep model to capture the  structures  particularly for food images from different aspects, the performance will be further improved. In addition, the constructed large-scale standard food dataset can also be critical to advance the development of  food recognition system.  There are more than 8,000 types of dishes worldwide according to Wikipedia \cite{Bola2017Food}. Compared with the large amount of dish types, the number of ingredients is limited. Therefore, one alternative solution is ingredient recognition. Some works \cite{Chen-DIRCRR-MM2016,Bola2017Food} have conducted multi-label ingredient prediction from  food images in terms of their lists of ingredients. Ingredient recognition will probably also a solution for offering an automatic mechanism for recognize images for applications in easing the tracking of the nutrition habits, leading to more accurate dietary  assessment.

\subsection{Joint Deep and Broad Learning for Food Computing}
A great amount of food-related data is being  recorded in various modalities, such as text, images and videos. It presents researchers with challenges, such as the sheer size of  data, the difficulty in understanding recipes, computer vision and  other machine learning challenges to study the culinary culture, eating habits and health. Fortunately, the recent breakthroughs in AI, especially the deep learning, provides powerful support for food data analysis from each data source. However, food related entities are from different networks, such as social networks, recipe-sharing websites and heterogeneous IoT sources. Effectively fusing these different information sources provides an opportunity for researchers  to understand the food data more comprehensively, which makes  ``Broad Learning'' an extremely important learning task. The aim of  broad learning is to investigate  principles, methodologies and algorithms to discover synergistic knowledge across multiple data sources \cite{Zhang2017BL}. Therefore, in order to learn, fuse and mine multiple food-related information sources with large volumes and multi-modality, one future direction is to jointly combine deep learning and broad learning from different data sources into a unified multimedia food data fusion framework. Such framework will provide a new paradigm, which is transformed to conventional food-related fields, such as food medicine and food science.

\subsection{Food-oriented Multimodal Knowledge Graph Construction and Inference}

We  can  exploit the enormous volume of food related data using sophisticated data analysis techniques to discover patterns and new knowledge. However, in order to support  heterogeneous modalities for more complex food-oriented retrieval, Question Answering (QA), reasoning and inference, a more effective method is to build a food-oriented multimodal knowledge graph incorporating visual, textual, structured data, rich context information,  as well as their diverse relations by  learning from large-scale multimodal food data. In natural language processing, some promising results have been shown  e.g., Freebase \cite{Bollacker2008Freebase}.  Semantic web technologies, e.g., ontologies and inference mechanism have been used for the diabete diet care \cite{Li2007Automated}. The study on  visual relationships with  triplets have been emerging in the area of computer vision, including the detection of visual relationships \cite{lu2016visual,Yaohui-DSL-AAAI2018} and generation of the scene graph \cite{Johnson2015Image} from images. These technologies are helpful for constructing the visual web \cite{Jain2015Let}. Other works such as \cite{Yuke-CoRR2015} tried to build a large-scale multimodal knowledge base system to support visual queries, and  have been shown as a promising way to construct the food-oriented multimodal knowledge graph. Such multimodal knowledge graph is useful to consistently represent the food data from various heterogeneous data sources. In addition, the reasoning can also be conducted based on the knowledge graph for supporting complex  query, QA and multimodal dialog via the inference engines.

\subsection{Food Computing for Personal Health}
Modern multimedia research has been fast developed in many fields such as art and entertainment, but lags in the health domain. Food is a fundamental element for the health. Food computing is  emerging as a promising field for the health domain, and can be used  to quantify the lifestyle and navigate the personal health. Recently, some works such as \cite{Nag-HML-ICMR2017,NagPSLWJ17} have proposed the life navigation system for future health ecosystems, such as the cybernetic health. \cite{Karkar2017TummyTrials} proposed a TummyTrials app, which can aid a person in  analyzing self-experiments to predict which type of  food can trigger their symptoms.  Food computing will provide  principles and methodologies for the integration and understanding of  food data produced by users. Combined with other information such as  attitudes and beliefs about food and recipes, the person's food preferences, lifestyles and hobbies, we can construct the personal model for personalized and health-aware food recommendation service. Therefore, one important direction is to apply food computing to build the personal model for the health domain.

\subsection{Food Computing for Human Behavior Understanding}
Earlier studies have demonstrated that the food affects the human behavior \cite{Kolata1982Food}. Different food choices lead to different change of behaviors. For example, food additives and unhealthy  diet could help to explain  criminal behavior alcoholism\footnote{https://articles.mercola.com/sites/articles/archive/2008/07/29/what-s-in-that-how-food-affects-your-behavior.aspx}. There are also some works  on  the relationship between food and human behavior, such as the eating behavior \cite{Tsubakida-Prediction-IJCAI2017,Achananuparp2018Does}, the brain activity \cite{Rosenbaum-LRWL-JCI2008} and cooking activities \cite{Stein-CEA-CPUC2013,Damen2018EPICKITCHENS}. For example, \cite{Ofli-Saki-WWW2017} utilized  large-scale food images from Instagram to study the food perception problem. \cite{Achananuparp2018Does} used the data from MyFitnessPal to analyze  healthy eating behaviors of users, who actively record food diaries. Food computing can effectively utilized food-oriented different signals, and thus will provide new methodologies and tools to advance the development in this direction.

\subsection{Food Log-oriented Food Computing}
 With the widespread use of mobile devices, e.g., digital cameras, smartphones and iPad,  people can easily take  photos of your food to record their diets. In addition, text-based meal record is also supported. Therefore, food logs records users' eating history with multimodal signals,  With the economic growth of the world, more and more people resorts to food logs for recording their general diet via the smartphone. Food log-oriented food computing will become more and more important for its multifarious applications. (1) Food logs are most critical for health.  Some works \cite{Waki-MFRT-DST2015}, \cite{Kitamura2008Food} \cite{Aizawa-FoodLog-IEEEMM2015} proposed a food-logging system, which is capable of distinguishing food images from other types of images for the analysis of  food balance. For example, \cite{Aizawa-FoodLog-IEEEMM2015} have proposed the FoodLog system\footnote{http://www.foodlog.jp/en}, which can receive access to all sorts of dietary information based on your sent photos by smartphones for the health management. In order to more precisely calculate daily intake of calorie from these multimodal signals, a robust food log oriented food recognition is also needed. (2) Food logs record  what  one eats or drinks daily and thus reflect their eating habits. Therefore, mining and analyzing rich food log data will enable personalized food recommendation,  which can offer healthier options for health-aware food recommendation \cite{Trattner-RCookingInterests-UMAP2017}. In addition, food logs record current popular food. We can  aggregate the food log data with  time stamps from millions of uses for food popularity prediction.

\subsection{Other Promising  Applications in the Vertical Industry}
There are also other promising applications for food computing in many vertical fields. For example, food computing  can enable many  applications in the smart home field, such as smart kitchen and personal nutrition log. Smart home systems can collect valuable information about users' preferences, nutrition intake and health data via food computing methods, such as food recognition and  cooking video understanding. Some existing  works, such as \cite{Kojima-AVSU-IROS2015} utilized the text information to understand the audio-visual scene  for a cooking support robot. In the future, we believe that the smart kitchen robot needs  more functions, more intelligent multimodal interaction and dialog. Food recognition, recipe recommendation and food-related text processing will work jointly to enable this goal. It will also play an important role in  the smart farming. Existing works such as \cite{Hern-SOC-ICTI2017,Chen2017Counting} can recognize and count the fruits  in the trees. More and more food computing systems will be applied to help detect the illness of the food to guarantee the food safety and quantity. With the development of food computing, it will also be applied into more emerging vertical fields, such as smart retails (especially for the grocery shopping) and smart restaurants.

\section{CONCLUSIONS}\label{section_CONCLU}
Food has a profound impact on many aspects of human, such as the survival, identity, religion and culture. Food computing can connect  food and human to improve  human health, understand human behaviors and culture.
The massive amount of food-related data  from various sources and the advances in  computer science and other principles (e.g., neuroscience and cognitive science) have provided us with unprecedented opportunities to tackle  many food-related issues via food computing. In this survey, we provide an extensive review of the most notable works to date on the datasets, tasks and applications of  food computing, from  food-oriented data acquisition and analysis, perception, recognition, retrieval, recommendation, prediction and monitoring to its various applications and services. This survey  discusses some key challenges in  food computing including robust and accurate food image recognition,vision based dietary management system, multiple-network oriented food data fusion and mining, health-aware personalized food recommendation and its applications for food science. Finally, this survey  suggested some research directions, such as large-scale standard food dataset construction, large-scale robust food recognition system, joint deep and broad learning for food computing, food-oriented multimodal knowledge graph construction and inference, food computing for food logs and other emerging vertical fields. These lines of promising directions need further research.

%

\bibliographystyle{ACM-Reference-Format}
\bibliography{food_csur_survey}

\end{document}